\documentclass[aps,prd,twocolumn,superscriptaddress]{revtex4-1}
\usepackage{graphicx}
\usepackage{amsmath,amssymb}
\usepackage{float}
\usepackage{bm}
\usepackage{color}

\begin{document}

\title{Thermal Broadening of Bottomonia: Lattice Non-Relativistic QCD with Extended Operators}

\author{Rasmus Larsen}
\affiliation{Physics Department, Brookhaven National Laboratory, Upton, New York 11973, USA}
\author{Stefan Meinel}
\affiliation{Department of Physics, University of Arizona, Tucson, Arizona 85721, USA}
\affiliation{RIKEN-BNL Research Center, Brookhaven National Laboratory, Upton, New York 11973, USA}
\author{Swagato Mukherjee}
\affiliation{Physics Department, Brookhaven National Laboratory, Upton, New York 11973, USA}
\author{Peter  Petreczky}
\affiliation{Physics Department, Brookhaven National Laboratory, Upton, New York 11973, USA}

\begin{abstract}

We present lattice non-relativistic QCD calculations of bottomonium correlation
functions at temperatures $T \simeq 150-350$~MeV. The correlation functions were
computed using extended bottomonium operators, and on background gauge-field
configurations for 2+1-flavor QCD having physical kaon and nearly-physical pion
masses.  We analyzed these correlation functions based on simple theoretically-motivated
parameterizations of the corresponding spectral functions.  The results of our
analyses are compatible with significant in-medium thermal broadening of  the ground
state S- and P-wave bottomonia.

\end{abstract}

\date{\today}
\maketitle

\section{Introduction}

At temperatures above the chiral cross-over temperature $T_c=156.5 \pm 1.5$
MeV~\cite{Bazavov:2018mes} hadrons start to melt and chiral symmetry gets restored.
However, quarkonia, bound states of charm or bottom quark and anti-quark can exist at
higher temperatures. It has been conjectured by Matsui and Satz that color screening
will eventually lead to melting of quarkonia, and suppression of quarkonium
production in heavy ion collisions can serve as signals of formation of deconfined
medium in such collisions~\cite{Matsui:1986dk}. The study of in-medium properties of
quarkonia and their production in heavy ion collisions is an extensive research
program, see Refs.~\cite{Aarts:2016hap,Mocsy:2013syh} for recent reviews. For
interpretation of experimental results on quarkonium production in heavy ion
collisions it is necessary to know quarkonium properties at non-zero temperature.

In-medium quarkonium properties are encoded in the spectral function,
$\rho(\omega,T)$, which can be related to the corresponding Euclidean correlation
function calculable on the lattice as
\begin{eqnarray}
&& C(\tau,T) = \int_0^{\infty} d\omega \rho(\omega,T) K(\tau,\omega,T) \, ,
\quad \mathrm{where} \nonumber \\
&& K(\tau,\omega,T) = \frac{\cosh(\omega(\tau-1/(2T)))}{\sinh(\omega/(2T))} \, .
\label{spec_rep}
\end{eqnarray}
Reconstructing the spectral function from discrete set of lattice data points for
$C(\tau,T)$ is a difficult task. Early attempts in this direction have been reported
in
Refs.~\cite{Nakahara:1999vy,Asakawa:2000tr,Asakawa:2003re,Wetzorke:2001dk,Karsch:2002wv,Datta:2002ck,Umeda:2002vr,Datta:2003ww,Jakovac:2006sf}.
It has been pointed out that Euclidean correlation functions have limited sensitivity
to the in-medium quarkonium properties and/or their
melting~\cite{Mocsy:2007yj,Petreczky:2008px}, because the maximal time extent is
limited to $\tau_\mathrm{max}=1/(2T)$ and becomes small at high temperature (c.f., Eq.
\ref{spec_rep}).

Studying bottomonium on the lattice is also challenging because of the large bottom
quark mass, $M_b \sim 5$ GeV, which implies large discretization effects $\sim a
M_b$, with $a$ being the lattice spacing. To circumvent this problem an effective
field theory approach can be used by integrating out the energy scale related to the
bottom quark mass. The corresponding theory is called the lattice non-relativistic QCD
(NRQCD)~\cite{Lepage:1992tx,Thacker:1990bm}. The heavy quark fields in this theory
are represented by Pauli spinors coupled to gauge fields, and quark--anti-quark pair
creation of the heavy quark is not allowed. Lattice NRQCD has been used to study bottomonium
properties at zero
temperature~\cite{Davies:1994mp,Meinel:2009rd,Meinel:2010pv,Hammant:2011bt,Dowdall:2011wh,Daldrop:2011aa,Lewis:2012ir,Wurtz:2015mqa}.
Attempts to calculate bottomonium spectral functions at non-zero temperatures using lattice
NRQCD  have also been
reported~\cite{Aarts:2010ek,Aarts:2011sm,Aarts:2012ka,Aarts:2013kaa,Aarts:2014cda,Kim:2014iga,Kim:2018yhk}.
Since pair creation is not allowed in NRQCD, the quark fields do not satisfy periodic
boundary condition~\cite{Brambilla:2008cx}. Therefore, the relation of the Euclidean
time correlator and the spectral function has the form~\cite{Aarts:2010ek}
\begin{equation}
C(\tau,T)=\int_{-\infty}^{+\infty} d \omega \rho(\omega,T) e^{-\tau \omega}\,.
\end{equation}
As a result, the maximal time extent at non-zero temperature is
$\tau_\mathrm{max}=1/T$, i.e., twice larger than in the relativistic case. For this
reason, the meson correlators in NRQCD are more sensitive to in-medium properties of
quarkonium~\cite{Aarts:2010ek,Aarts:2014cda,Kim:2014iga,Kim:2018yhk}.

Till now, all the lattice NRQCD studies at non-zero temperature have used point-like
meson operators for computations of the quarkonium correlation functions.
The use of point-sources for the non-relativistic quarkonia is appealing, since
the corresponding spectral function has the physical interpretation in terms of di-lepton rate, and
the melting of quarkonia is easily understood as disappearance of peaks in the corresponding spectral functions.
However, it is well-known that point-like hadron operators, usually, do not have good
overlap with a particular meson state in that quantum number channel, and precise
results $C(\tau,T)$ for large $\tau$ are needed to isolate the contribution of a
particular meson to the correlation function. As a result, point meson operators are
not very sensitive to in-medium modifications of quarkonia.

Extended operators are widely used to study quarkonium properties at zero temperature
\cite{Davies:1994mp,Meinel:2009rd,Meinel:2010pv,Hammant:2011bt,Dowdall:2011wh,Daldrop:2011aa,Lewis:2012ir,Wurtz:2015mqa}.
Correlators of extended operators have a better projection onto quarkonium states and
are less sensitive to the continuum, i.e., the bound-state contribution to these
correlators is significantly larger. For this reason, we expect that correlators of
extended meson operators will be more sensitive to in-medium quarkonium modifications
and/or their dissolution. In this paper we explore the temperature dependence of
bottomonium correlators corresponding to extended meson sources within lattice NRQCD.
As we will see later, compared to the correlators of point meson operators, the
correlators of extended sources have significantly larger temperature dependence. We
try to understand the observed temperature dependence of the correlation functions of
extended meson operator using simple theoretically-motivated parameterization for the
in-medium spectral functions, and find evidence for thermal broadening of bottomonium
states.

The rest of the paper is organized as follows. In section II we describe our setup of
lattice calcualtions. We present our results at zero temperature in section III.
Non-zero temperature case and bottomonium properties in the deconfined medium are
discussed in section IV. Section V contains our conclusions.

\section{Details of lattice QCD calcualtions}

In NRQCD the heavy quarks and anti-quarks are described in terms of two component
Pauli spinors, $\psi$ and $\chi$. In our study, we use tree-level tadpole-improved
NRQCD action. The tadpole improvement means that the gauge-link variables that enter
the NRQCD Lagrangian on the lattice are divided by the average value of the link,
$u_0$. The NRQCD action used in this study includes all terms of order $v^4$ as well
as $v^6$ spin-dependent terms, with $v$ being the heavy-quark velocity inside
quarkonium.

The calculations in NRQCD are set-up as an initial value problem. The heavy-quark
propagator,
\begin{equation}
G_{\psi}(\mathbf{x},t) = \left\langle \psi(\mathbf{x},t) \: \psi^\dag(\mathbf{0},0) \right\rangle \,,
\label{g_psi}
\end{equation}
is calculated as follows
\begin{eqnarray}
G_{\psi}(t) &=& K(t) G_{\psi}(t-1) \,,
\quad \mathrm{with} \nonumber \\
K(t) & =&  \left( 1-\frac{a \delta H|_t}{2} \right) \left(1-\frac{a H_0 |_t}{2n}\right)^n U_4^\dagger (t)
\nonumber \\ [2mm]
 &\times & \left(1-\frac{a H_0 | _{t-1}}{2n}\right)^n \left( 1-\frac{a \delta H|_{t-1}}{2} \right) \,.
\label{evol}
\end{eqnarray}
Here,  $t=\tau/a$ and $U_4(t)$ is the temporal link. The $H_0=-\Delta ^{(2)}/(2 M_b)$
is the leading order non-relativistic Hamiltonian,  and
\begin{eqnarray}
   \Delta ^{(2)}q(\mathbf{x},t) &=& \frac{1}{u_0}\sum _{j=1} ^3
\left[ U_j(\mathbf{x},t)q(\mathbf{x}+\hat{j},t) \right. \nonumber \\
&+& \left. U_{j}^{\dagger}(\mathbf{x},t)q(\mathbf{x}-\hat{j},t) \right]
-  6 q(\mathbf{x},t) \,.
\end{eqnarray}
is the covariant lattice Laplacian. Here, $\delta H$ is the part of the NRQCD
Lagrangian that contains corrections of order $v^4$ and spin dependent $v^6$
correction. The explicit form of $\delta H$ is given in Ref. \cite{Meinel:2010pv}.
The parameter $n$ plays the role of the discretization time step and is called the
stability parameter or the Lepage parameter. For the tadpole parameter $u_0$ we use
the forth root of the plaquette. The anti-quark propagator
\begin{equation}
G_\chi(\mathbf{x},t) = \left\langle \chi(\mathbf{0},0) \: \chi^\dag(\mathbf{x},t) \right\rangle.
\label{g_chi}
\end{equation}
is also determined by Eq. (\ref{evol}), since it
can be related to quark propagator as $ G_\chi(\mathbf{x},t) = - G_\psi^\dag(\mathbf{x},t)$.
In the above equations we use the same notation as in Ref. \cite{Meinel:2010pv}.

In this paper we are mostly interested in the  correlators of extended meson operators of the form
$O=\tilde \chi^{\dagger} \Gamma \tilde \psi$,
obtained from smeared quark and anti-quark fields,
\begin{eqnarray}
  & \widetilde{\psi} = W \psi \,,
  \quad \mathrm{and} \quad
  \widetilde{\chi} = W \chi \,,
  \quad \mathrm{with} \quad \nonumber \\
  & W = \left(1 + \frac{\sigma^2}{4N}\Delta^{(2)}\right)^{N}.
\end{eqnarray}
For sufficiently large number of the smearing steps, $N$, the meson operator has a
Gaussian shape having a width $\sigma$. The root mean square size of the meson source
is $r_{RMS}=\sqrt{3} \sigma/2$. To avoid instabilities, one has to choose $N > 3
\sigma^2/2$. Spin and color structures of the meson are determined by the matrix
$\Gamma$. Along with the corresponding lattice irreducible representations (irrep),
the choices for $\Gamma$ used in the present study of bottomonia are given in
Table~\ref{tab:Gamma}, in terms of the Pauli matrices $\sigma_i$ and the covariant
derivative operator
\begin{equation}
  \nabla_i q(\mathbf{x},t) = \frac{1}{u_0} \left[
  U_i(x,t)q(\mathbf{x}+\hat{i},t) - U_i^{\dagger}(\mathbf{x},t)q(\mathbf{x}-\hat{i},t)
  \right] \,. \nonumber
\end{equation}

\begin{table}[!t]
\begin{center}
 \begin{tabular}{ccl}
 \hline\hline
 Name & Irrep $\Lambda^{PC}$ & $\Gamma$ \\
 \hline
 $\eta_b$    & $A_1^{-+}$         & 1\!\!1 \\
 $\Upsilon$  & $T_{1}^{--}$       & $\sigma_j$ \\
 $h_b$      & $T_{1}^{+-}$       & $\nabla_j$ \\
 $\chi_{b0}$ & $A_1^{++}$         & $\boldsymbol{\sigma}\cdot\boldsymbol{\nabla}$ \\
 $\chi_{b1}$ & $T_1^{++}$         & $(\boldsymbol{\sigma}\times\boldsymbol{\nabla})_j$ \\
 $\chi_{b2}$ & $T_2^{++}$         & $\sigma_j\nabla_k + \sigma_k\nabla_j\hspace{3ex}(j\neq k)$ \\
\hline\hline
\end{tabular}
 \caption{\label{tab:Gamma} Structures used in the interpolating fields.}
 \end{center}
\end{table}

We are interested in the meson two-point function, which can be expressed in terms of the quark propagator as
\begin{eqnarray}
  C(t) &=& \sum_{\mathbf{x}} \left\langle O(\mathbf{x},t) O^\dag(\mathbf{0},0) \right\rangle \nonumber \\
  &=& - \sum_{\mathbf{x}} \mathrm{Tr}\left\{ \left[W G_\psi W^\dag\right]^\dag\!(\mathbf{x},t,\mathbf{0},0)
  \right. \nonumber \\
  &\times&   \left[\Gamma W G_\psi W^\dag \Gamma^\dag\right] \!(\mathbf{x},t,\mathbf{0},0)
  \left. \phantom{\left[W G_\psi W^\dag\right]^\dag} \hspace{-1.9cm}\right\} \, .
  \label{meson}
\end{eqnarray}
In the above equations the gauge links entering the smearing operator $W$ and its conjugate
are defined on time-slice indicated in parenthesis, i.e. the gauge link entering $W$ are
defined on time slice $t$, while the gauge link entering $W^{\dagger}$ are defined on time-slice zero.

We choose $N$ and $\sigma$ such that $r_{RMS}$ is about $0.21$ fm for all lattice
spacings. This choice provides a good overlap with the ground state for both S-state and P-state bottomonium correlators at T=0. We use this value also at finite temperature. At finite temperature the physical size of the states might have changed. We therefore explored the effect of $r_{RMS}$ at finite temperature in appendix \ref{app:size_dep}. To make contact with earlier
studies we also calculated the correlators of point meson operators that are obtained
from Eq. (\ref{meson}) by setting $W$ to unit matrix.

\begin{table*}[!t]
\begin{center}
  %\begin{tabular}{ | l | c | c | c | c | c | c |}
  \begin{tabular}{  l  |  c   c  c  c  c  c }
    \hline \hline
    $\beta $ & 6.740 & 6.880 &  7.030 & 7.280 & 7.373 & 7.596 \\[2mm]
    %\hline
    a[fm] & $0.1088(2)$ & $0.0951(2)$ &  $0.0825(2)$ &  $0.0655(2)$ & $0.0602(2)$  & $0.0493(2)$ \\[2mm]
    %\hline
    $u_0$ & 0.87288 & 0.87736 & 0.88173 & 0.88817 & 0.89035 &0.89517 \\[2mm]
    %\hline
    $n$ (Lepage) & 4 &  4 & 4 & 4 & 4 & 8 \\[2mm]
    %\hline
    $\sigma$  & 2.19089 & 2.50679 & 2.88929 & 3.63923 & 3.95980 &  4.83487 \\[2mm]
    %\hline
    $N$       & 10  & 13    & 17    & 27     &  33     &  49\\[2mm]
    %\hline
    size $T=0$ & $48^4$ & $48^4$ &  $48^4$ & $48^3 \times 64$ & $48^3 \times 64$ & $64^4$ \\[2mm]
    %\hline
    $\#$ conf $T=0$ & 192 & 192 &  192 & 384 &  512 & 384 \\[2mm]
    %\hline
    $\#$ Sources $T=0$ & 32 & 32 &  32 & 32 &  8 & 16 \\[2mm]
    %\hline
    $T$ [MeV]  & 151.1 & 172.9 &  199.3 & 251.0 & 273.1 & 333.5\\[2mm]
    %\hline
    size $T>0$ &  $48^3 \times 12$ & $48^3 \times 12$ & $48^3 \times 12$ & $48^3 \times 12$ & $48^3 \times 12$ & $48^3 \times 12$ \\[2mm]
    %\hline
    $\#$ conf $T>0$ & 384 & 768  &  768 & 1152 & 384 & 768 \\[2mm]
    %\hline
    $\#$ sources $T>0$ & 32 & 32  &  32 & 32 & 64 & 32 \\[2mm]
    %\hline
    $t_{min}$ Tuning & 6 & 8  & 8 & 11 & 11 & 16 \\[2mm]
    %\hline
    $t_{max}$ Tuning & 23 & 23  & 23 & 31 & 31 & 31 \\[2mm]
    %\hline
    $aM_b$ & $2.341(9)$ & $2.047(8)$  & $1.745(6)$& $1.351(6)$ & $1.22(1)$ &$0.957(9)$ \\[2mm]
    %\hline
$a M_{\eta_b}$ & $0.4560(3) $ & $0.4511(4)$  & $0.4550(4)$& $0.4958(2)$ & $0.532(1)$ &$0.6245(2)$ \\
    \hline \hline
  \end{tabular}
\end{center}
\caption{Parameters of lattice calculations, including the lattice sizes, the lattice
spacings, the number of gauge configurations, the number of sources per gauge
configurations, the tadpole parameters, the values of the Lepage (stability)
parameter, and the values of $\lambda$ and $N$ for Gaussian smearing. The fit ranges
of the meson correlators, the tuned NRQCD mass parameters and the energies of the
$\eta_b$ state are also presented in the last four rows.
}
\label{tab:param}
\end{table*}

In our study we use 2+1 flavor gauge configurations generated by the HotQCD
collaboration using highly improved staggered quark (HISQ) action at the physical
strange quark mass and light quark masses corresponding to the pion mass of $161$ MeV
in the continuum limit \cite{Bazavov:2011nk,Bazavov:2014pvz}. Thus, our calculations
are performed almost at the physical point. We perform calculations on $48^3 \times
12$ lattices in the temperature range $T=151-334$ MeV. These correspond to lattice
spacings $a=0.05-0.11$ fm. For each temperature we perform the corresponding
calculations at $T=0$.  We use multiple sources when evaluating meson correlation
functions. The parameters of the lattice calculations, including  the lattice sizes,
lattice spacings number of gauge configurations, number of sources per gauge
configurations, tadpole parameters and the values of the Lepage (stability) parameter
are given in Tab.~\ref{tab:param}.

For a given value of the heavy quark mass parameter, $M_b$, we determine the masses
of quarkonium states by fitting the large $\tau$ behavior of the meson correlation
functions with a single exponential. As already mentioned above, correlators of
extended meson operators have a better projection onto the lowest state in a given
channel.
In order to judge to what extent correlators are dominated by the lowest
energy state we consider  the effective mass, defined as
\begin{equation}
  M_{eff}(\tau,T) = \frac{1}{a}\log \left[ \frac{C\left(\tau,T\right)}{C\left(\tau+a,T\right)} \right] \,.
\end{equation}

\begin{figure}[!t]
  \centering
  \includegraphics[width=0.45\textwidth]{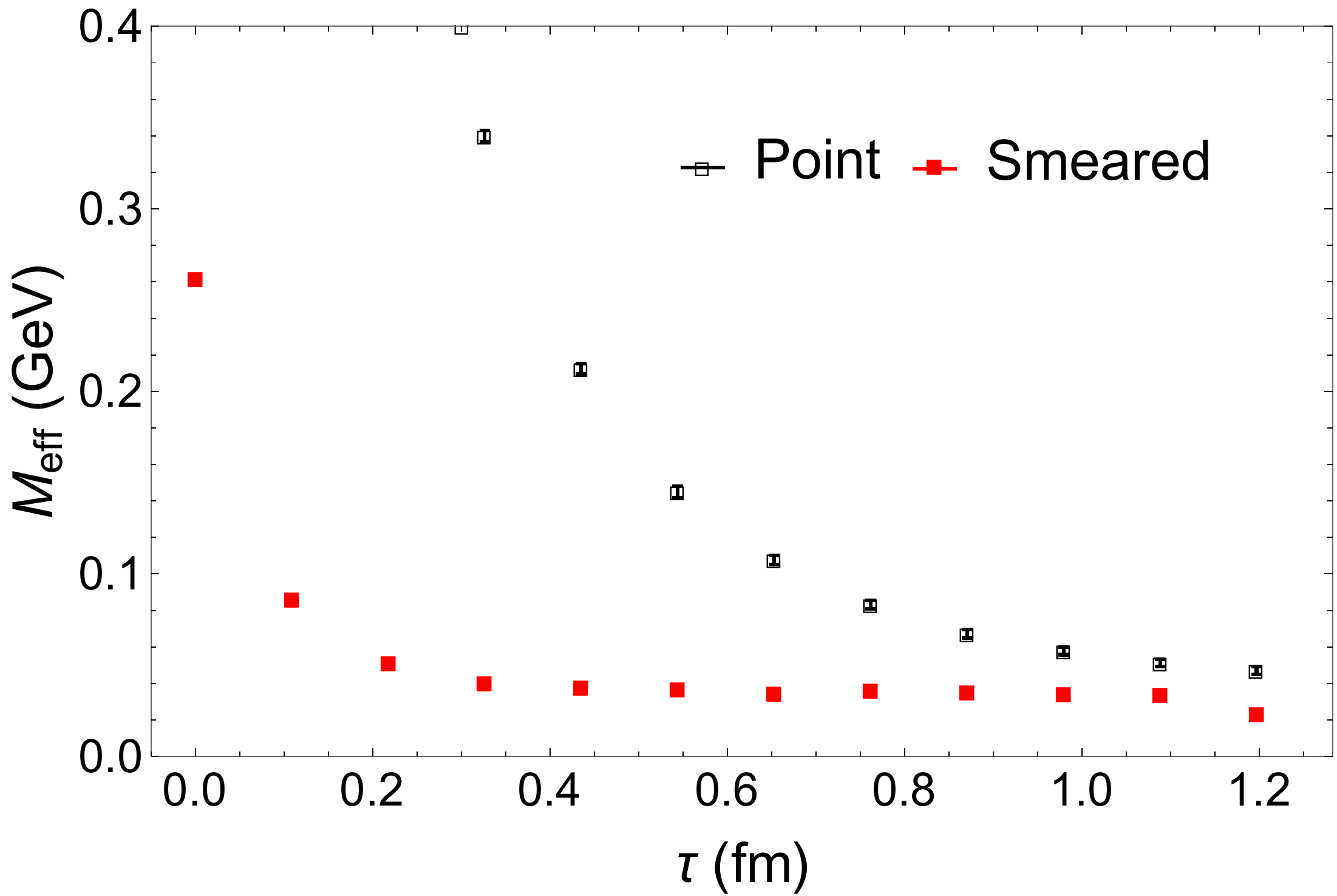}
   \caption{Effective mass for $\Upsilon$ at $T=0$ using point (black) and smeared
   (red) sources. The vertical scale is calibrated with the $T=0$ $\eta_b$ mass.}
   \label{fig:smear_vs_point}
\end{figure}

In Fig.~\ref{fig:smear_vs_point} we show the effective masses from $\Upsilon$
correlators at $T=0$ for extended (smeared) and point sources. For point sources the
effective mass reaches a plateau only for $\tau>1.2$~fm, while for the extended
sources the plateau is reached already for $\tau \simeq 0.4$~fm, implying that the
correlation function is dominated by the ground state contribution at relatively
small $\tau$. This will be important for the analysis of the correlation functions at
$T>0$, where the time extend is limited by the inverse temperature. The highest
temperature used in our study is $T=333.5$~MeV, which corresponds to $\tau_{max}
\simeq 0.6$~fm. Thus, we expect to  probe quarkonium properties even at the highest
temperature  using correlators with extended sources. To obtain the $T=0$ bottomonium
energy levels we fit correlators with smeared sources  to a single exponential form
in the interval $[t_{min}:t_{max}]$. The value of $t_{min}$ is chosen such that it
corresponds to $\tau_{min}>0.5$~fm to eliminate the possibility of excited state
contamination. The value of $t_{max}$ is determined by the signal-to-noise ratio of
the correlators and the desire to have well controlled statistical errors. In
Tab.~\ref{tab:param} we give the values of $t_{min}$ and $t_{max}$ use in the fits.

\section{Numerical results at zero temperature}

\begin{figure}[!t]
  \centering
  \includegraphics[width=0.45\textwidth]{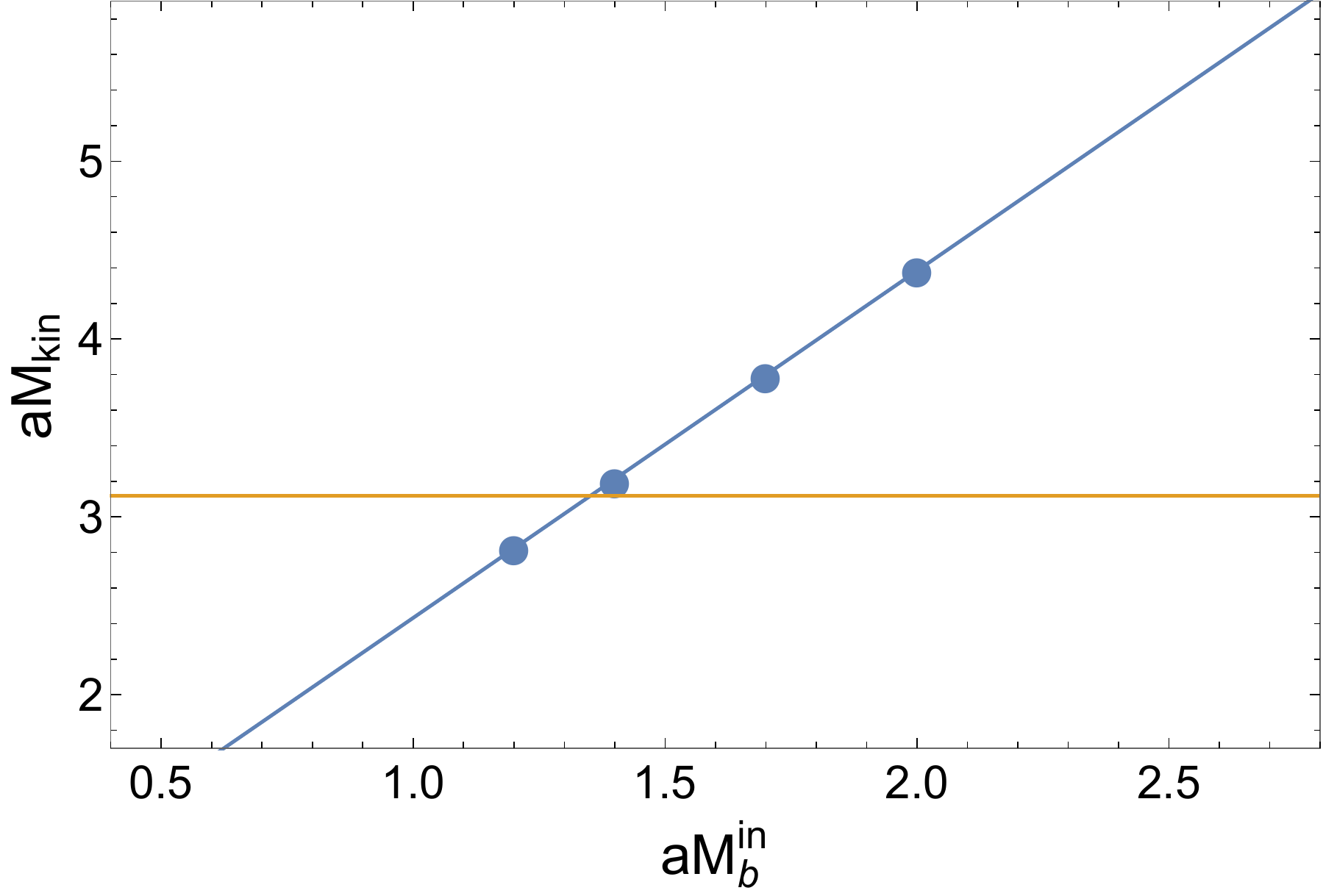}
   \caption{Kinetic mass of $\eta _b$, in lattice units, as a function of  $a M_b^{in}$ for $a=0.0655$~fm. The horizontal line
    corresponds to the PDG mass.}
   \label{fig:Mkin}
\end{figure}

In this section we discuss our results on bottomonium masses and correlators at $T=0$
\footnote{strictly speaking we do not have results at exactly zero temperature
because of the finite extent of the temporal direction. However, for large enough
temporal extent as used here thermal effects are not visible within our present
statistical accuracy.}, which provide an essential baseline for the study of
bottomonium properties in the medium. Before we can study the bottomonium properties we
have to fix the mass parameter, $M_b$, in the NRQCD Lagrangian. This can be done by
studying the dependence of the energy of the bottomonium state as function of the
spatial momentum $\mathbf{p}$,
\begin{equation}
E(p)=\sqrt{\mathbf{p}^2+ M_{kin}^2}+const \,,
\label{Mkin}
\end{equation}
defining the so-called kinetic mass, $M_{kin}$, of the meson~\cite{Meinel:2010pv}. In
this study we use Eq.~(\ref{Mkin}) for the kinetic mass, however, the
non-relativistic definition gave very similar numerical results. We determine the
kinetic mass of $\eta_b$ meson for a given input mass parameter $M_b^{in}$. Then we
interpolate the kinetic mass of the $\eta_b$  in $M_b^{in}$ and determine the
physical value of the mass parameter by requiring that $M_{kin}=M_{\eta_b}^{PDG}$,
$M_{\eta_b}^{PDG}$ being the experimental value of the $\eta_b$ mass from Particle
Data Group (PDG)~\cite{PDG18}. This procedure is demonstrated in Fig.~\ref{fig:Mkin}.
In practice, we find that linear interpolation in $M_b^{in}$ works well. The values
of the tuned $M_b$ for different $\beta$ are given in Tab.~\ref{tab:param}.

NRQCD is expected to break down for $a M_b<1$, because radiative corrections become
very large when the inverse lattice spacing is larger than the mass parameter, see
e.g. Ref.~\cite{Dowdall:2011wh}. In our setup this happens for lattice spacing of
about $0.04$~fm. This breakdown can be seen in the behavior of the kinetic mass as
function of the mass parameter $aM_b$. The kinetic mass decreases monotonically with
decreasing $aM_b$. This decrease is usually well described by a linear dependence,
c.f. Fig.~\ref{fig:Mkin}. For sufficiently small $aM_b$ we start seeing deviations
from the linear behavior, and eventually the kinetic mass does not decrease with
decreasing $aM_b$, but starts to saturate at some value.
% In other words, the kinetic mass of the meson cannot be made smaller than some value for a given
% lattice spacing by decreasing $M_b$.
We see that for $\beta=7.825$ corresponding to $a=0.04$~fm we cannot reach the
physical $\eta_b$ mass by lowering $aM_b$. For this reason, $\beta=7.596$ was the
largest gauge coupling used in our study. For charmonia the largest $\beta$ value
that can be used is $\beta=6.74$. Therefore, in-medium properties of charmonia cannot
be studied within the present NRQCD framework and using HotQCD lattices. Attempts to
study charmonia at non-zero temperature using NRQCD have been presented in
Ref.~\cite{Kim:2018yhk}. In that work the NRQCD parameter was fixed to be
$M_c=1.275$~GeV for all values of $\beta$. Our analysis shows that this choice of the
mass parameter results in $\eta_c$ mass above $4$~GeV for $\beta>6.74$.

Having determined $M_b$ we can study the spectrum of bottomonium states at $T=0$. In
NRQCD absolute values of the meson masses cannot be calculated, only the
corresponding energy levels that are related to the masses by a lattice spacing
dependent constant can be determined. Therefore, we will consider the differences
between the energy levels of various quarkonium states and the energy levels of
$\eta_b$. These are equal to the corresponding differences in meson masses. In what
follows, we will use the energy of $\eta_b$ states to calibrate the energy scale at
different lattice spacings, i.e., we will set the energy of $\eta_b$ state to be
zero. With this in mind we will use the term mass and energy level interchangeably.
In Table~\ref{tab:mass} we show the difference of the masses of various bottomonium
states and the $\eta_b$ mass. These differences are compared to the experimental
values from PDG~\cite{PDG18}. For P-wave bottomonia, namely for $\chi_{b0}$,
$\chi_{b1}$, $\chi_{b1}$ and $h_b$ the mass difference agrees well with the
experimental results. The difference between the $\Upsilon$ and the $\eta_b$ mass,
i.e., the hyper-fine splitting is smaller than the experimental value. This is
similar to the findings of Ref.~\cite{Meinel:2010pv}, where order $v^6$ NRQCD action
was used. One needs the radiative corrections in the NRQCD Lagrangian in order to
reproduce the hyper-fine splitting~\cite{Dowdall:2011wh}.

Before studying bottomonium properties at non-zero temperature we would like to
understand the spectrum of energy levels encoded in the extended meson operators used
in this study. On general grounds, we expect that the spectrum of energy levels
consist of a ground state, one or two excited states below the open bottom
threshold, and many higher lying states that in the infinite volume limit form a
continuum. In the case of finite volume, the spectral function is always given by the
sum of $\delta$-functions. However, even in this case the density of states at large
$\omega$ is very large \cite{Kim:2018yhk} and, in practice, it is possible to approximate the spectral
function by a continuum. Thus, the spectral function can be approximated  as
\begin{equation}
\rho(\omega)=\sum_i A_i \delta(\omega-M_i)+\theta(\omega-s_0) \rho_{cont}(\omega).
\end{equation}
Here, $s_0$ is the open beauty threshold. For point sources
\begin{equation}
\rho_{cont} \sim (\omega-s_0)^{n},
\label{cont}
\end{equation}
with $n=1/2$ for S-wave quarkonia and $n=3/2$ for P-wave quarkonia. For large values
of $\omega \sim 1/a$ the spectral function is distorted by lattice artifacts and
vanishes above some $\omega_{max}$ \cite{Kim:2018yhk}. For extended sources the form
of $\rho_{cont}$ is not known, but the above general form of the spectral function is
still valid. Since only a few lattice data points in the correlation function are
sensitive to the high energy part of the spectral function, and excited state
contribution is suppressed when extended sources are used, we can employ a simplified
parameterization of the spectral function
\begin{equation}
\rho(\omega)=A \delta(\omega-M)+ \rho_{high}(\omega),
\label{spfT0}
\end{equation}

i.e., we can consider only the ground state contribution and some continuum contribution, which
has support only for $\omega>M$. The explicit form of $\rho_{high}$ is not important for
our analysis. If the above equation holds, the correlation function can be written as
\begin{eqnarray}
C(\tau) &=& A e^{-M \tau}+ C_{high}(\tau)\nonumber\\
C_{high}(\tau) &=& \int_{-\infty}^{\infty} d \omega \rho_{high}(\omega) e^{-\omega \tau}.
\label{CT0}
\end{eqnarray}
As discussed in the previous section, the correlators of extended operators are
dominated by the ground state for $\tau>0.5$~fm and, therefore, it is possible to
determine the parameters $A$ and $M$ from the single exponential fit in this $\tau$
region. From this we can obtain $C_{high}(\tau)$. In Appendix~\ref{ap_subtracted_cor}
we show the determination of $C_{high}(\tau)$ for different lattice spacing, which
will be used in the following sections to analyze the correlators at non-zero
temperature. In the initial tuning, we fitted with a ground state and a continuum at the same time. As seen in appendix \ref{app:size_dep} in Fig. \ref{etab_smooth_com} and \ref{pwave_smooth_com}, we see that the continuum is well explained by a simple step function.

\begin{table*}[!t]
 \begin{center}
  \begin{tabular}{  l | c  c  c  c  c  c  c  c }
  %\begin{tabular}{ | l | c | c | c | c | c | c | c | c |}
    \hline \hline
    $a[fm]$ & 0.1088 & 0.0951 &  0.0825  & 0.0655 & 0.0493 & PDG \\ [2mm] %\hline
    $\Upsilon$   & $35.3(1.5)$ & $36.5(1.0)$ & $38.6(1.6)$  & $42.1(1.0)$ & $51.2(1.6)$ & 61.3(2.3) \\[2mm] %\hline
    $\chi _{b0}$ & $460(14)$ & $456(7)$ & $459(7)$  & $460(5)$ & $459(11)$ & 460.5(2.4) \\[2mm] %\hline
    $\chi _{b1}$ & $477(11)$ & $478(8)$ & $481(7)$ & $486(6)$ & $487(11)$ & 493.8(2.3) \\[2mm] %\hline
    $\chi _{b2}$ & $489(9)$ & $491(8)$  & $499(9)$  & $503(7)$  & $505(11)$ & 513.21(2.3) \\[2mm] %\hline
    $h_b$ & $479(7)$ & $482(8)$  & $488(7)$  & $491(6)$  & $495(11)$ & 500.3(2.4) \\ \hline \hline
\end{tabular}
\caption{The mass differences of bottomonium states with respect to $\eta _b$ mass, in MeV.}
\label{tab:mass}
\end{center}
\end{table*}

\section{Finite Temperature Results}

The temperature range which we explore in this paper goes from $151$~MeV up to
$334$~MeV. As discussed in the previous section, the temperature range is limited by
the fact that the NRQCD approximation starts to break down for $aM_b< 1$. The last
temperature we look at is right on the boundary of this limit. We have explored all
temperatures with both point to point correlators, and smeared to smeared
correlators. The analyses of the $T>0$ correlators are performed in terms of the
effective mass, $M_{eff}$, calibrated with respect to the $T=0$ mass of $\eta_b$
meson.

\subsection{Correlators with point sources at $T>0$}

\begin{figure}[!t]
  \centering
  \includegraphics[width=0.45\textwidth]{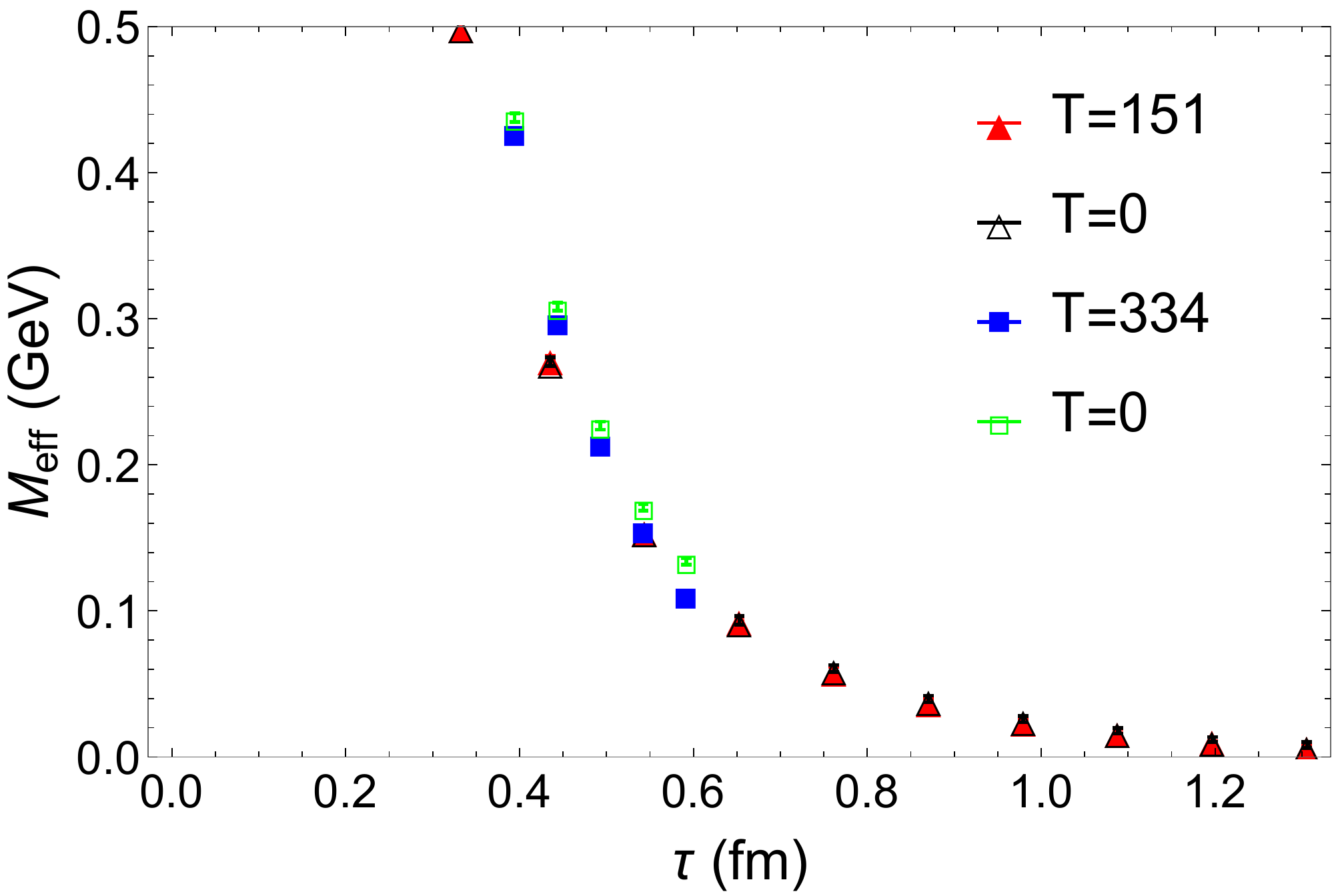}
   \caption{Effective mass for $\eta _b$ from point sources for $T=151.1$~MeV and
   $T=333.5$~MeV compared to the corresponding effective masses at zero temperature.
   The vertical scale is calibrated with the $T=0$ $\eta_b$ mass.}
   \label{fig:etab_point}
\end{figure}

\begin{figure}[!t]
  \centering
  \includegraphics[width=0.45\textwidth]{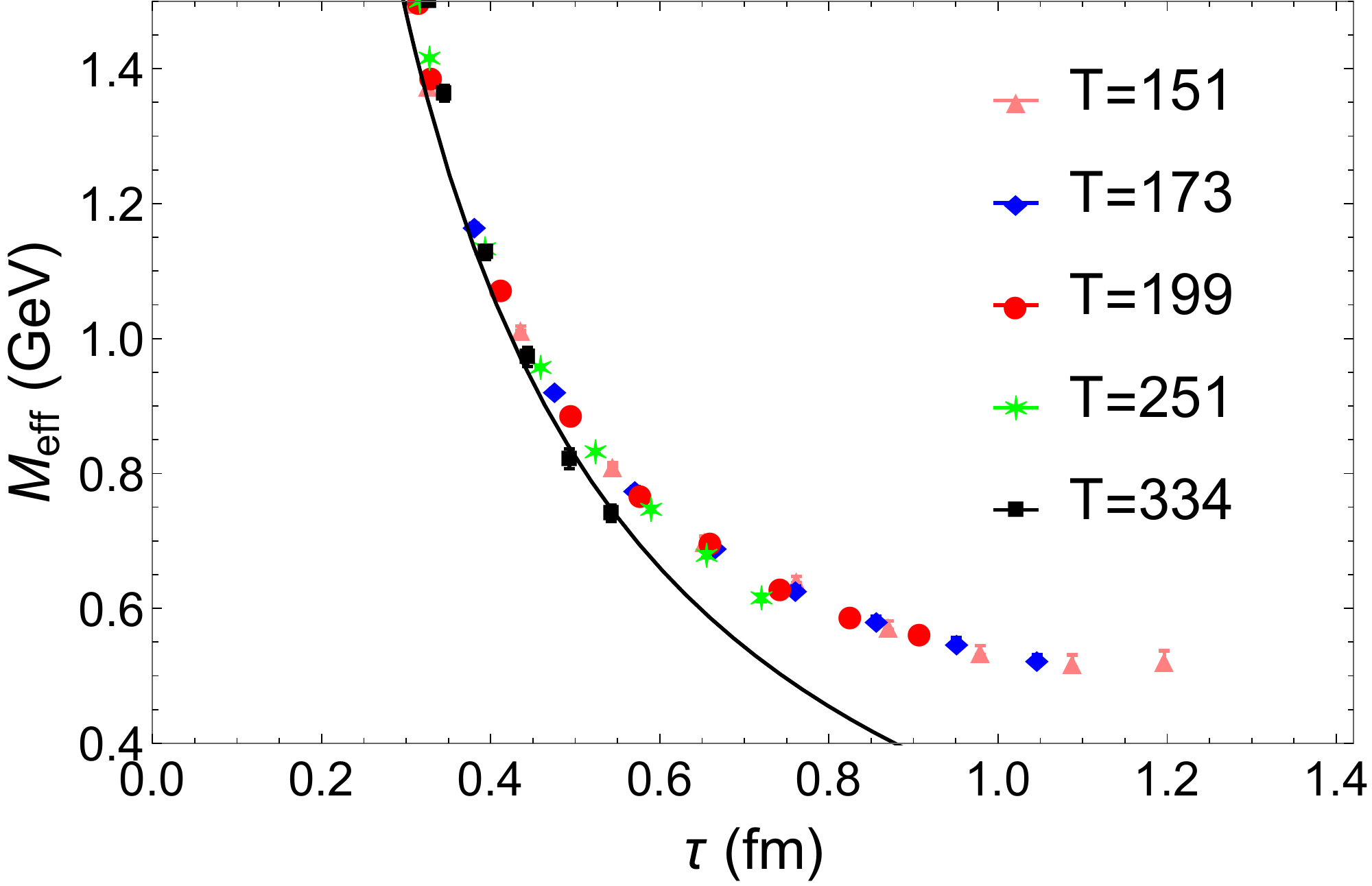}
   \caption{Effective mass for $\chi _{b0}$ from point sources minus the zero temperature $\eta_b$ mass
    for $T=151.1$~MeV (pink), $172.9$~MeV (blue), $199.3$~MeV (red), $251.0$~MeV(green) and $333.5$~MeV(black).
    The solid black line shows the effective mass corresponding to free quarks at $T=333.5$~MeV. The vertical scale is calibrated with the $T=0$ $\eta_b$ mass.}
   \label{fig:chib0_point}
\end{figure}

We start our discussion with the results obtained with point sources. In
Fig.~\ref{fig:etab_point} and Fig.~\ref{fig:chib0_point} we show the effective masses
corresponding to $\eta_b$ and $\chi_{b0}$ correlators, respectively. For the $\eta_b$
correlator we see very little temperature dependence. In fact, no medium effect can
be seen at the lowest temperature. At the highest temperature, T=334 MeV, we see a small but visible temperature dependence of the $\eta_b$ correlator. Somewhat larger temperature dependence is observed
for the $\chi_{b0}$ correlators. We see that the effective masses become  smaller
compared to the $T=0$ effective masses with increasing temperature and $\tau$. This
implies enhancement of the finite temperature correlators compared to the
corresponding one at zero temperature with increasing temperature and Euclidean time,
which is consistent with the previous studies of correlators with point
sources~\cite{Aarts:2014cda,Kim:2014iga,Kim:2018yhk}.

At sufficiently high temperatures we expect that bottomonium states will be dissolved
and the correlator, as well as the spectral function will be reasonably well described
by the free theory. This will happen at smaller temperature for the P-wave bottomonia
because of their larger size. In absence of lattice artifacts the free correlator of
the P-states will be proportional to $\tau^{5/2}$, c.f. Eq.~(\ref{cont}). Therefore,
the effective masses will have the form $M_{eff}=b+5/(2 \tau)$.
In Fig.~\ref{fig:chib0_point} we compare this expectations with the lattice results
for the $\chi_{b0}$ effective masses at the highest temperature, $T=333.5$~MeV.  The
constant was fixed to match with the lattice results at the largest available $\tau$. 
 
  The free-theory form of $M_{eff}$ does not
describe the lattice results for $\tau<0.4$ at a quantitative level.
Because of cutoff effects it is difficult to establish that at high temperatures the effective masses for $\chi_{b0}$ approach the free theory value. Based on this, we can therefore not claim to find that the results approach the free theory. We do however find that the effective mass is very similar to the results of Ref.~\cite{Aarts:2010ek}.

From the results shown in Fig.~\ref{fig:etab_point} and \ref{fig:chib0_point} we see
that the effective masses corresponding to correlators with point sources are
dominated by the large $\omega$ part of the spectral function, since if the low $\omega$ part was dominating, the T=0 effective mass would have reached a plateau in Fig. \ref{fig:etab_point} and \ref{fig:chib0_point}. This is not the case. Instead we see a strong falloff in the effective mass. Similar conclusion has been reached in
Ref.~\cite{Kim:2018yhk}. Therefore, in the following we will study in-medium
bottomonium properties using smeared Gaussian sources and sinks.

\subsection{$T \neq 0$ with smeared sources}

\begin{figure*}[!t]
  \centering
  \includegraphics[width=0.45\textwidth]{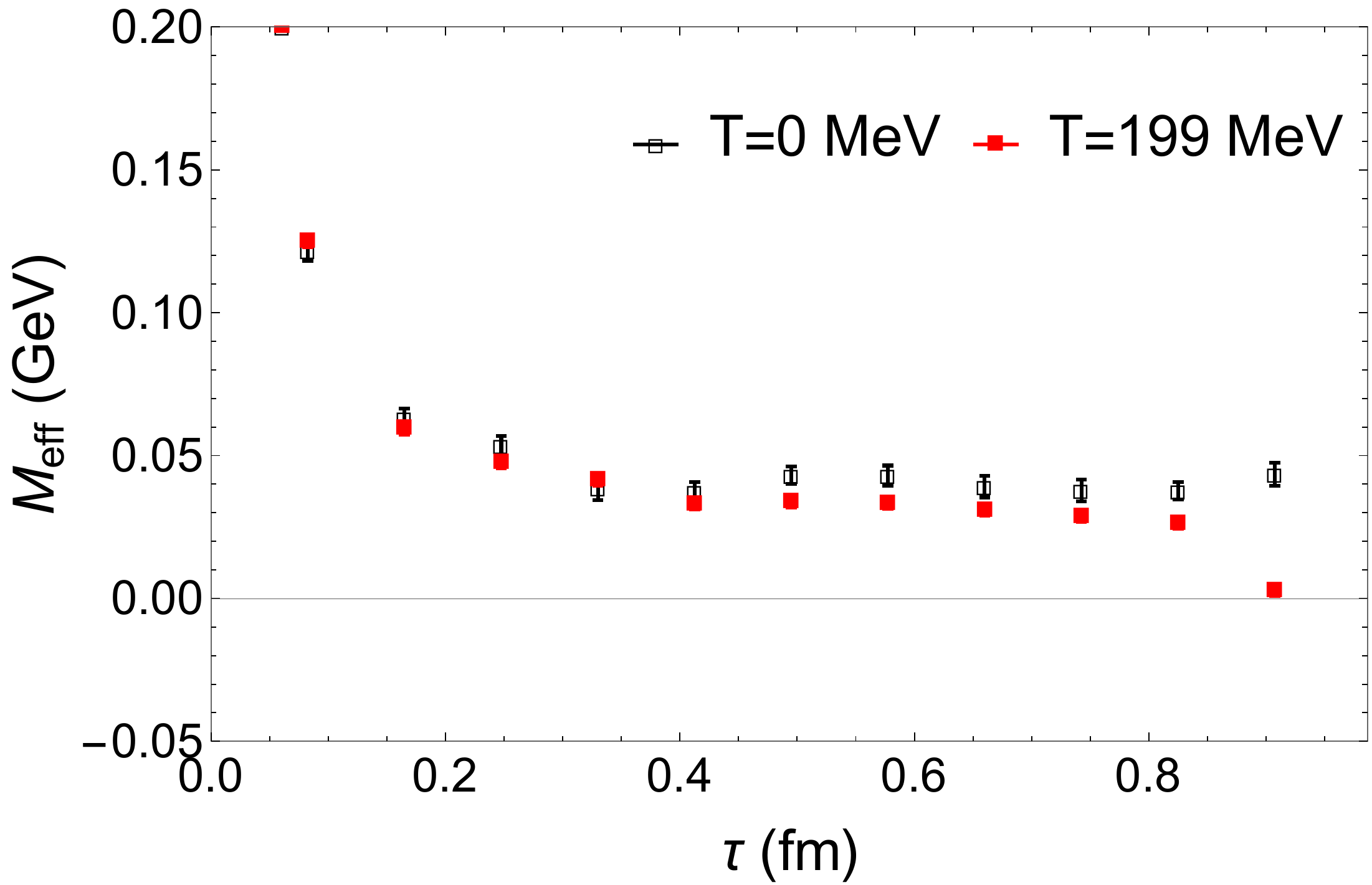}
  \hspace{0.05\textwidth}
  \includegraphics[width=0.45\textwidth]{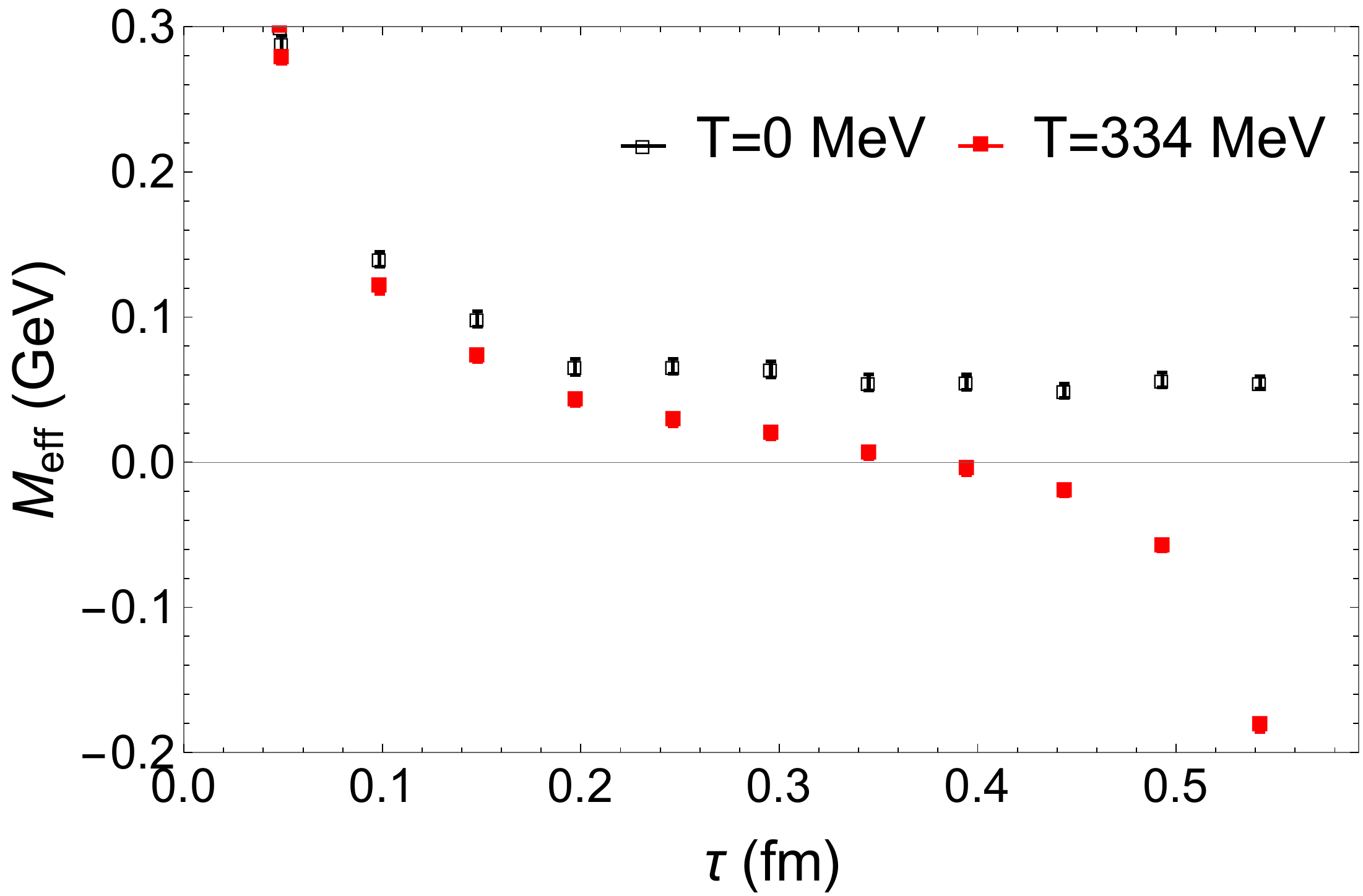}
  \\ \vspace{0.02\textheight}
  \includegraphics[width=0.45\textwidth]{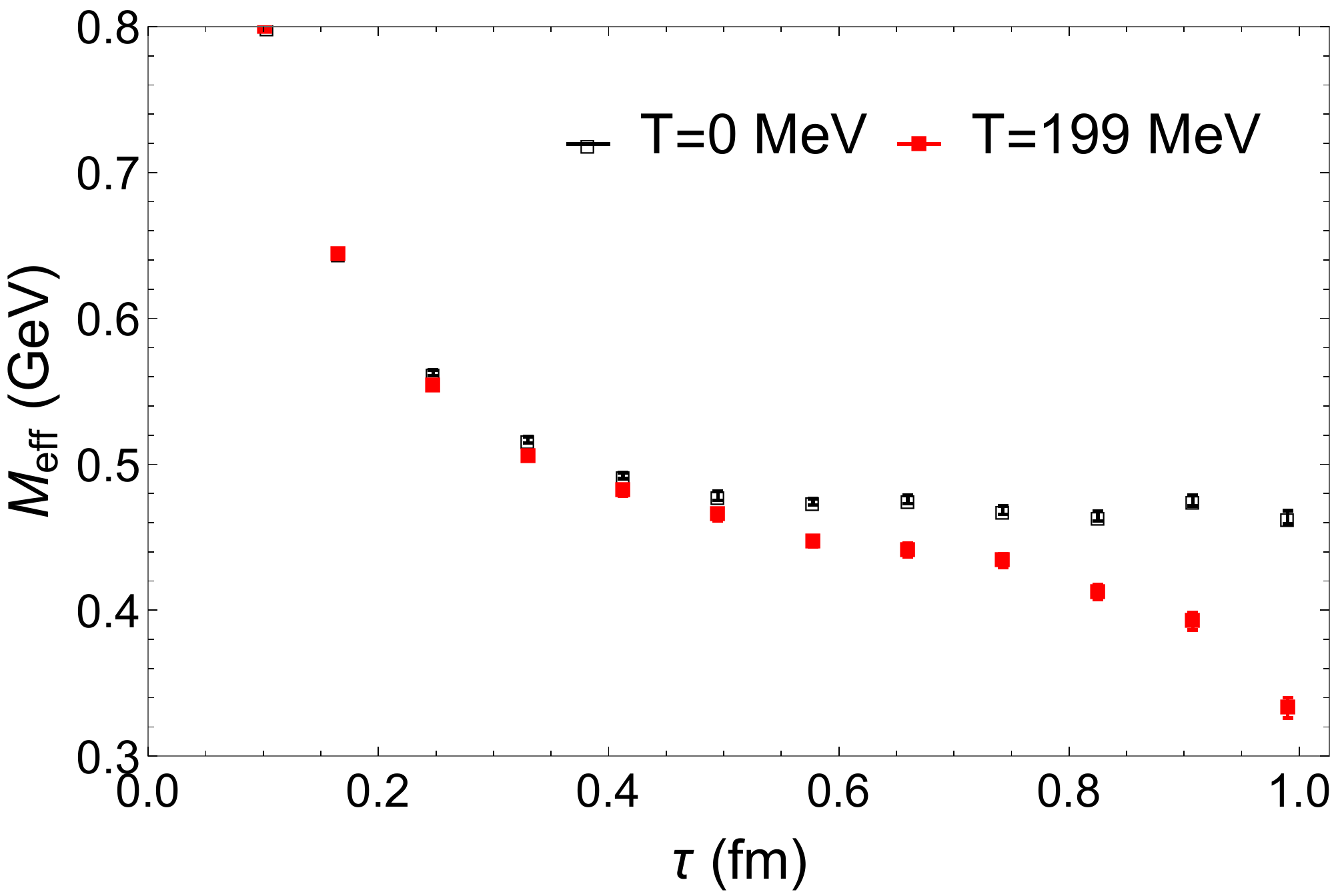}
  \hspace{0.05\textwidth}
  \includegraphics[width=0.45\textwidth]{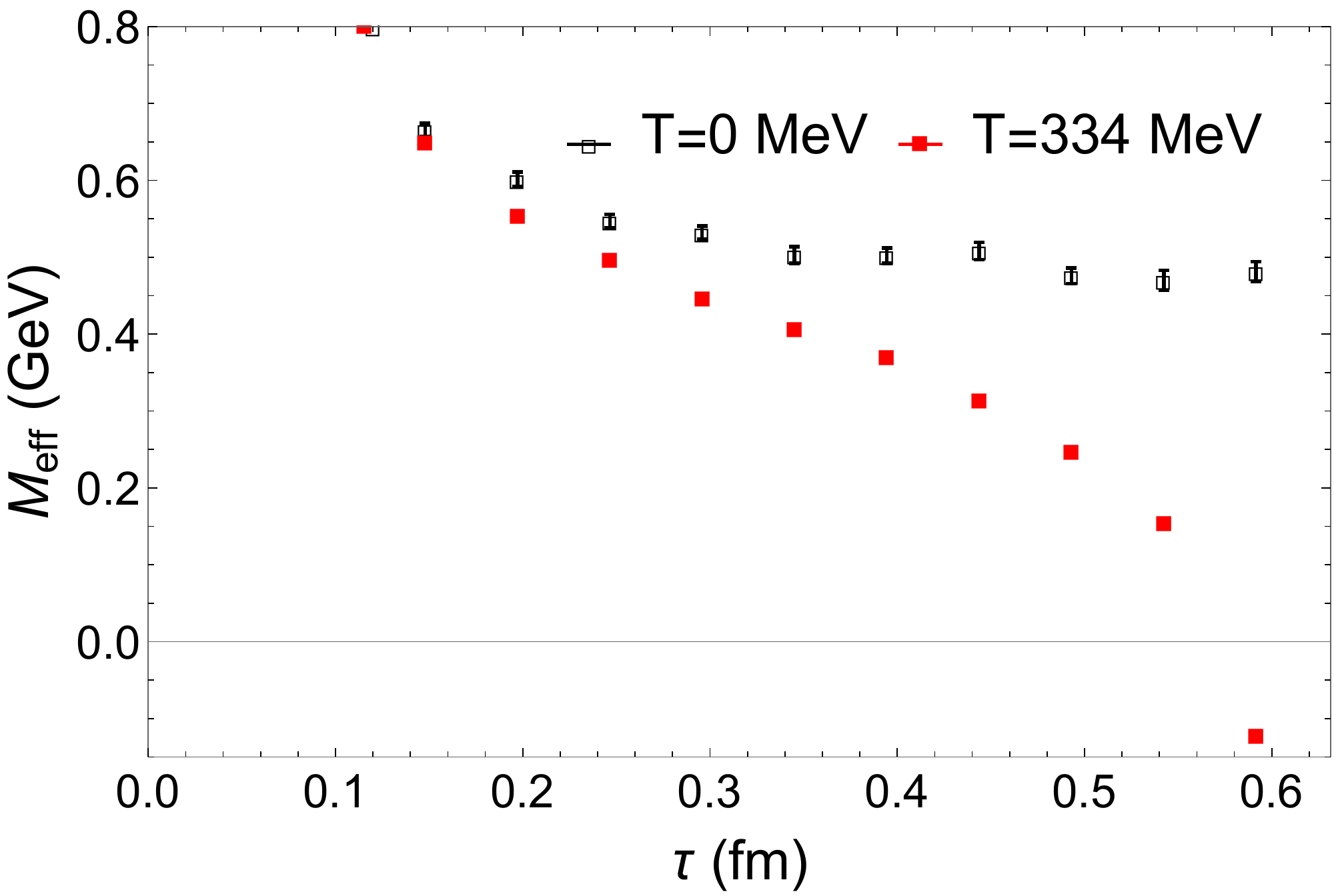}
   \caption{Effective masses for $\Upsilon$ (top panels) and $\chi_{b0}$ (bottom panels) from smeared correlators
   at two temperatures and $T=199$~MeV (left) and $333.5$~MeV (right). The results at zero temperature
   are also shown for comparison.}
   \label{fig:meff_ext_T}
\end{figure*}

We study the effective masses for the smeared correlators to get some insight about
in-medium modification of the bottomonium states. Our results for the effective
masses of $\Upsilon$ and $\chi_{b0}$ at two representative temperatures are shown in
Fig.~\ref{fig:meff_ext_T}. In the figure we also show the corresponding zero
temperature results for the reference. At small $\tau$ we see little to no
temperature dependence in the effective masses. For the lower temperature,
$T=199$~MeV, we see an approximate plateau in the $\Upsilon$ effective masses for
$\tau<0.8$~fm and a rapid drop at larger $\tau$. For the highest temperature the
$\Upsilon$ effective masses show a mild approximately linear decrease in the $\tau$
region, where the corresponding $T=0$ effective mass has an approximate plateau. At
large Euclidean time, $\tau>0.45$~fm we see again a rapid  drop in the effective mass
of the $\Upsilon$ correlator. The effective mass of $\chi_{b0}$ correlator at $T=199$
MeV shows a behavior that is very similar to that of the $\Upsilon$ effective mass at
the highest temperature. This is expected, since $\chi_{b0}$ state being larger is
more affected by the medium. The effective mass of $\chi_{b0}$ at the highest
temperature does not show any remnant of a plateau, but a significant decrease with a
slope that is increasing with increasing $\tau$.

A similar behavior in the effective masses to the one observed above has also been
found in the calculations of static meson correlation functions at finite
temperature, for sufficiently large separation between the static quark $Q$ and
static anti-quark $\bar Q$~\cite{Bazavov:2014kva,Petreczky:2017aiz}. In the case of
static mesons, the temperature and $\tau$ dependence of the effective masses can be
more easily understood. For sufficiently high temperatures, the  spectral function of
a static meson can be calculated in Hard Thermal Loop (HTL) re-summed perturbation
theory~\cite{Burnier:2013fca}. Perturbative calculations show that the energy of
static $Q\bar Q$ pair is complex~\cite{Laine:2006ns,Brambilla:2008cx}. This energy is
also often referred to as the complex $Q\bar Q$ potential. Therefore, the spectral
function of a static meson has a peak at small $\omega$. The position of the peak is
given by the real part of the potential, while the imaginary part of the potential
determines the width of the peak. We also expect that there is a continuum part of
the spectral function at large $\omega$~\cite{Rothkopf:2011db}. Around the peak
position, the shape of the spectral function is well described by a Lorentzian (or
skewed Lorentzian)~\cite{Burnier:2013fca}. However, the peak has also long tail at
small $\omega$~\cite{Burnier:2013fca}, which is not related to the imaginary part of
the potential~\footnote{We thank Y. Burnier for discussions on this point.}. This
tail determines the large $\tau$ behavior of the correlation function of static
$Q\bar Q$ meson. The qualitative features of the spectral function of static $Q\bar
Q$ meson obtained in HTL re-summed perturbation theory can also help to explain the
behavior of the effective masses of static meson correlators observed in lattice
calculations~\cite{Bazavov:2014kva,Petreczky:2017aiz}. In particular, it can be shown
that the large $\tau$ behavior of the effective masses is determined by tails of the
spectral function at small $\omega$.

\begin{figure*}[!t]
  \centering
  \includegraphics[width=0.45\textwidth]{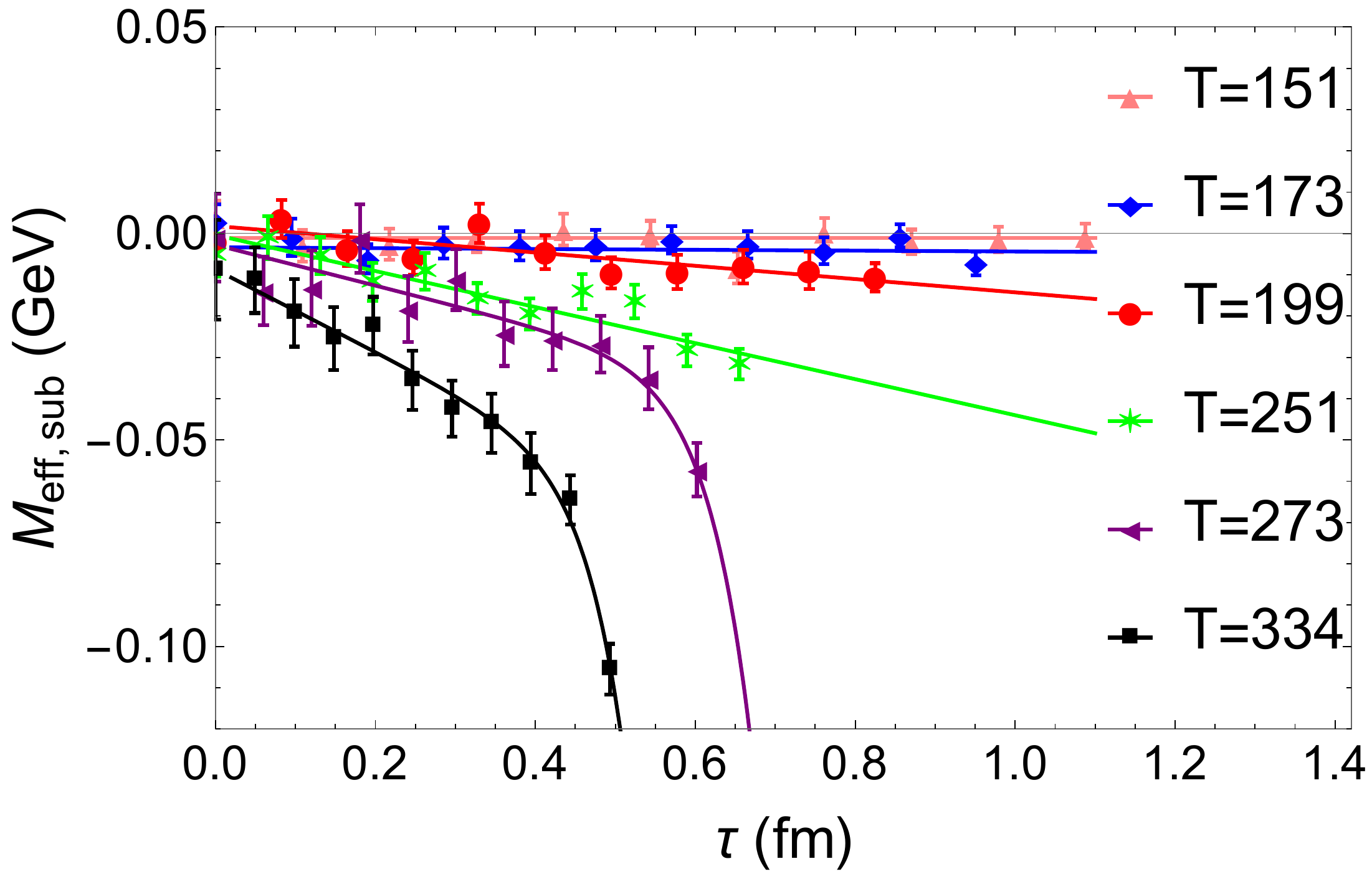}
  \hspace{0.05\textwidth}
  \includegraphics[width=0.45\textwidth]{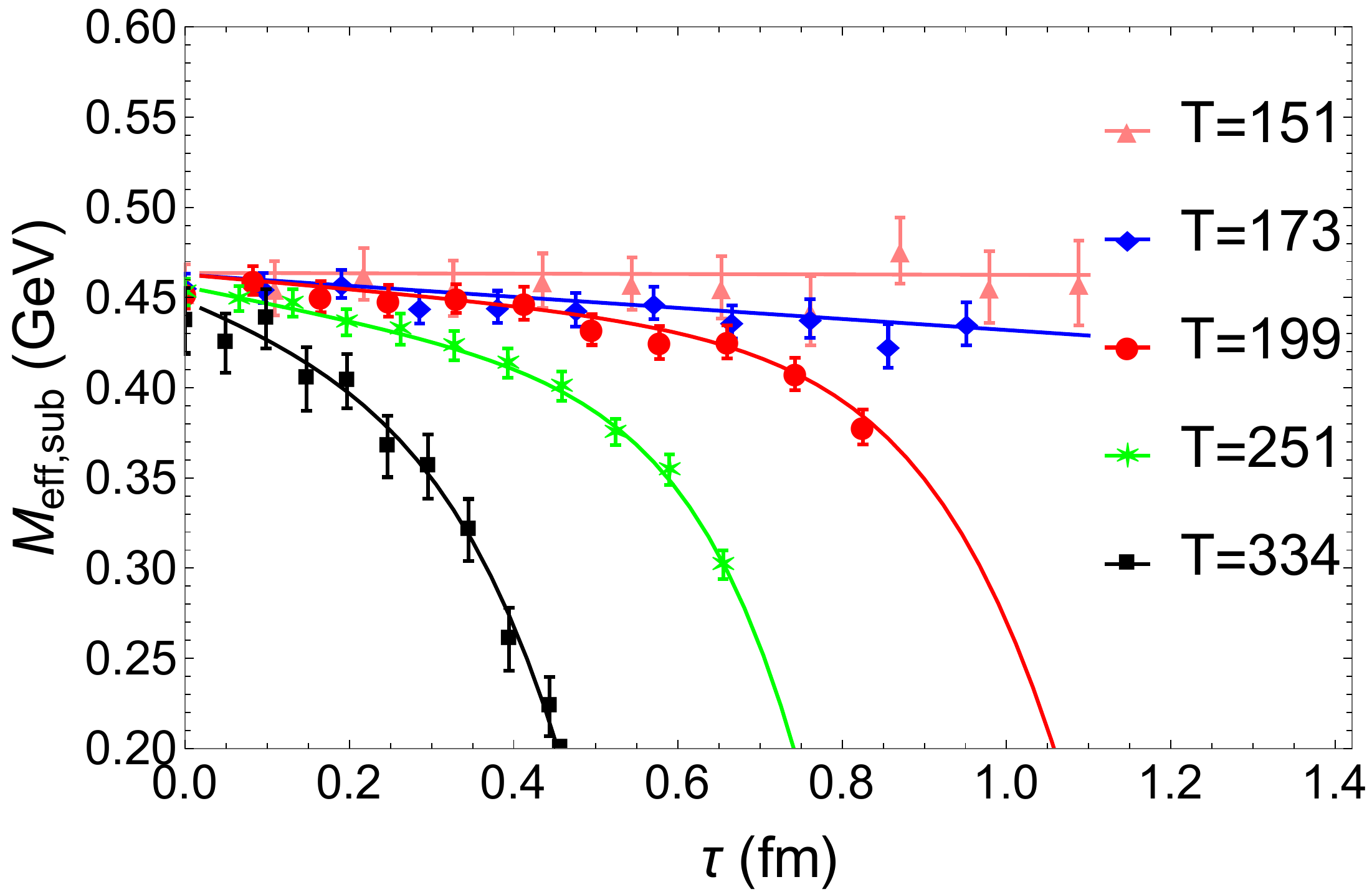}
   \caption{Zero temperature subtracted effective masses for $\eta_b$ (left) and
   $\chi_{b0}$ (right) correlators as function of $\tau$ at different temperatures:
   $T=151.1$~MeV (pink), $172.9$~MeV (blue), $199.3$~MeV (red), $251.0$~MeV (green),
   $273.1$~MeV (purple) and $333.5$~MeV (black). The lines correspond to fits using
   Eq.~(\ref{gauss}).}
   \label{fig:dmeff_fit_gauss}
\end{figure*}

Given the above discussion it is reasonable to assume that the spectral function at
$T>0$ has a broadened peak around $\omega$ that corresponds to a quarkonium state,
and a high frequency part that is identical to the zero temperature one, i.e., we can
assume that the spectral function has the form
\begin{equation}
\rho(\omega,T)=\rho_{med}(\omega,T)+\rho_{high}(\omega) \,.
\label{rhoT}
\end{equation}
Here, $\rho_{med}(\omega,T)$ describes the spectral functions for $\omega \simeq M$
and/or $\omega < M$. A natural parameterization of $\rho_{med}(\omega,T)$ would be a
Breit-Wigner (Lorentzian) form
\begin{equation}
\rho_{med}(\omega,T)=\frac{1}{\pi} \frac{\Gamma}{(\omega-M)^2+\Gamma^2} \,,
\end{equation}
with $M$ and $\Gamma$ being the temperature dependent mass and width of the
bottomonium state, respectively.  The Lorentzian form is expected to capture well the
main features of the spectral function for $\omega \sim M$, but also has a long tail
for $\omega \ll M$, where it is not adequate. As discussed above in the case of
static mesons, the Lorentzian only works in the vicinity of the peak and the
functional dependence on $\omega$ is very different from the Lorentzian for $\omega$
well below the peak position. Since we do
not know the functional form of the spectral function at very small $\omega$, we
assume that
\begin{equation}
\rho_{med}(\omega,T)=\frac{1}{\pi} \frac{\Gamma}{(\omega-M)^2+\Gamma^2} \theta(\omega-\omega_{cut}) \,.
\label{eq_cut_lorentzian}
\end{equation}
The lorentzian form parameterize the tail of the spectral function, but need to be cutoff, due to the falloff being too slow. It turns out that the Euclidean correlation
function and the effective masses are sensitive to the choice of $\omega_{cut}$ at large
$\tau$. Thus, we need at least three parameters to describe the medium dependent part
of the spectral function, $\rho_{med}$: the peak position, $M$, the thermal width
$\Gamma$, and the $\omega_{cut}$ that parameterize the tail of the spectral function
at small $\omega$.

The above discussion holds in the infinite volume limit. In lattice calculations the
volume is finite and, therefore, the number of available energy levels is also
finite. Therefore, $\rho_{med}(\omega,T)$ should be given by a sum of delta
functions. Furthermore, for typical volumes used in present day lattice calculations
the number of energy levels is not very large for $\omega \sim M$, see discussions in
Ref.~\cite{Kim:2018yhk}. In particular, we should not expect that there are many
delta functions in the lattice volume that will effectively parameterize the low
$\omega$ tail of the spectral function.

At small $\tau$ the correlators and the effective masses are mostly sensitive to the
high energy part of the spectral function. Since the high energy part of the spectral
function is temperature independent, the effective masses at small Euclidean time are
not sensitive to the effects of the medium on quarkonium states. One should consider
only large $\tau$ values to gain sensitivity to in-medium bottomonium properties. On
the other hand, as discussed above, at large $\tau$ the behavior of the correlators
is sensitive to the small $\omega$ tail of the spectral function, which is also
unrelated to bottomonium properties at $T>0$. Therefore, we consider the subtracted
correlator
\begin{equation}
C_{sub}(\tau,T)=C(\tau,T)-C_{high}(\tau).
\end{equation}
If Eq.~(\ref{rhoT}) is valid, the subtracted correlator at small $\tau$ should be
sensitive to $\rho_{med}(\omega,T)$ at $\omega \simeq M$.

We can also define an effective mass corresponding to the subtracted correlator
$M_{eff,sub}(\tau,T)$, which is shown in Fig.~\ref{fig:dmeff_fit_gauss} for $\eta_b$
and $\chi_{b0}$. There are three key features of $M_{eff,sub}(\tau,T)$: at small
$\tau$ it is close to $T=0$ bottomonium mass, at intermediate $\tau$ we see a mild,
approximately linear decrease, and finally, there is a sharp drop at large $\tau$
values. The fact that $M_{eff,sub}(\tau,T)$ is close to the $T=0$ mass at small
$\tau$ may suggest that the quarkonium mass is not significantly shifted relative to
its vacuum value. The mild linear decrease in the difference of the effective masses
is closely related to the width of the bottomonium state. To extract in-medium
bottomonium properties we fit the data on the subtracted effective masses using the
following Ansatz for $\rho_{med}$
\begin{eqnarray}
\rho_{med}(\omega,T) &=& A_{cut} \delta(\omega-\omega_{cut}) \nonumber \\
&+& A \exp \left[ -\frac{ \left( \omega-M(T) \right)^2}{2 \Gamma^2(T)} \right] \,.
\label{gauss}
\end{eqnarray}
The Gaussian form was chosen, since its falloff is sufficiently fast, and gives a linear dependence only, in the effective mass. The parameters $A_{cut}$ and $\omega_{cut}$ effectively describe the tail of the spectral function at low $\omega$.
If $A_{cut}$ is small, the subtracted effective mass corresponding to the above equation has the form
$M_{eff,sub}=M(T)-\Gamma^2 \tau$, i.e., the Gaussian form naturally explains
the approximate linear dependence of the effective masses. At large $\tau$ the tail of the spectral function
at small $\omega$ also becomes important, and we see deviations from
the linear behavior.

\begin{figure*}[!t]
\centering
\includegraphics[width=0.45\textwidth]{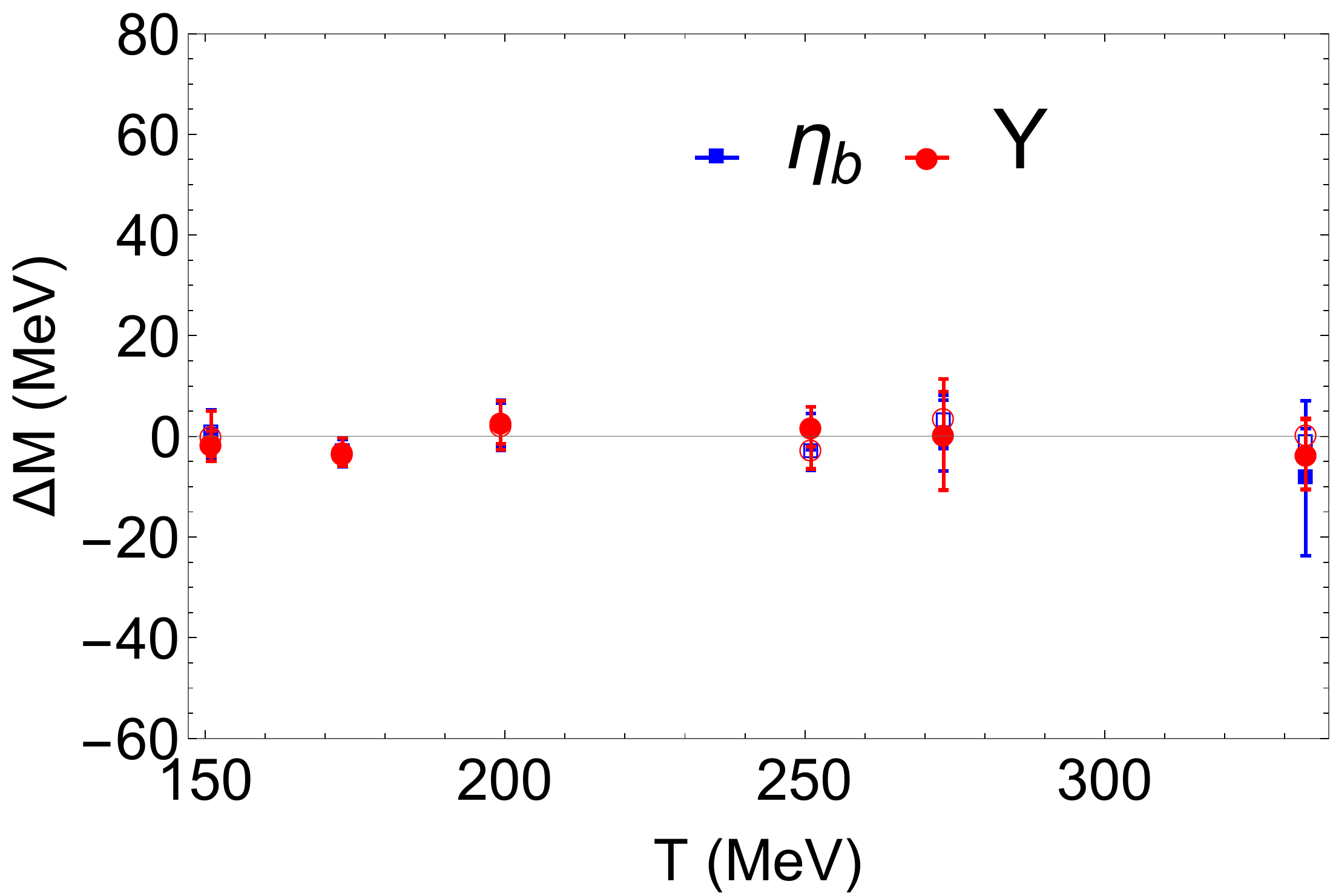}
\hspace{0.05\textwidth}
\includegraphics[width=0.45\textwidth]{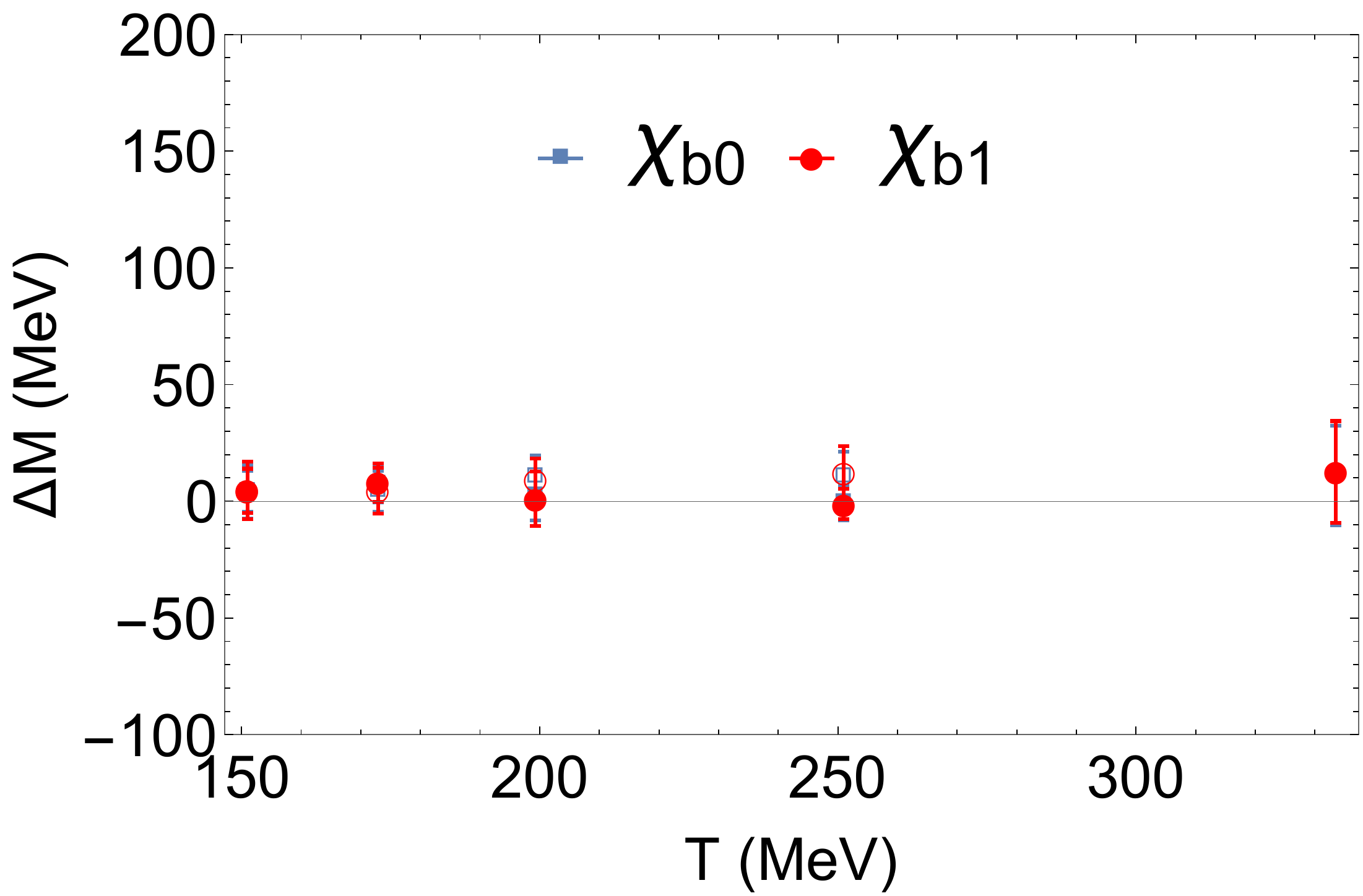}
\caption{In-medium mass shifts from fits to Eq.~(\ref{gauss}) for 1S bottomonium
(left) and 1P bottomonium (right). The open symbols correspond to fits that exclude
the data at the largest two $\tau$ values and $A_{cut}=0$, while the filled symbols
correspond to fits with all the data points included. Open points does not work for
$\chi _b$ at the highest temperature and are thus not shown.}
\label{fig:massT}
\end{figure*}

\begin{figure*} [!t]
\centering
\includegraphics[width=0.45\textwidth]{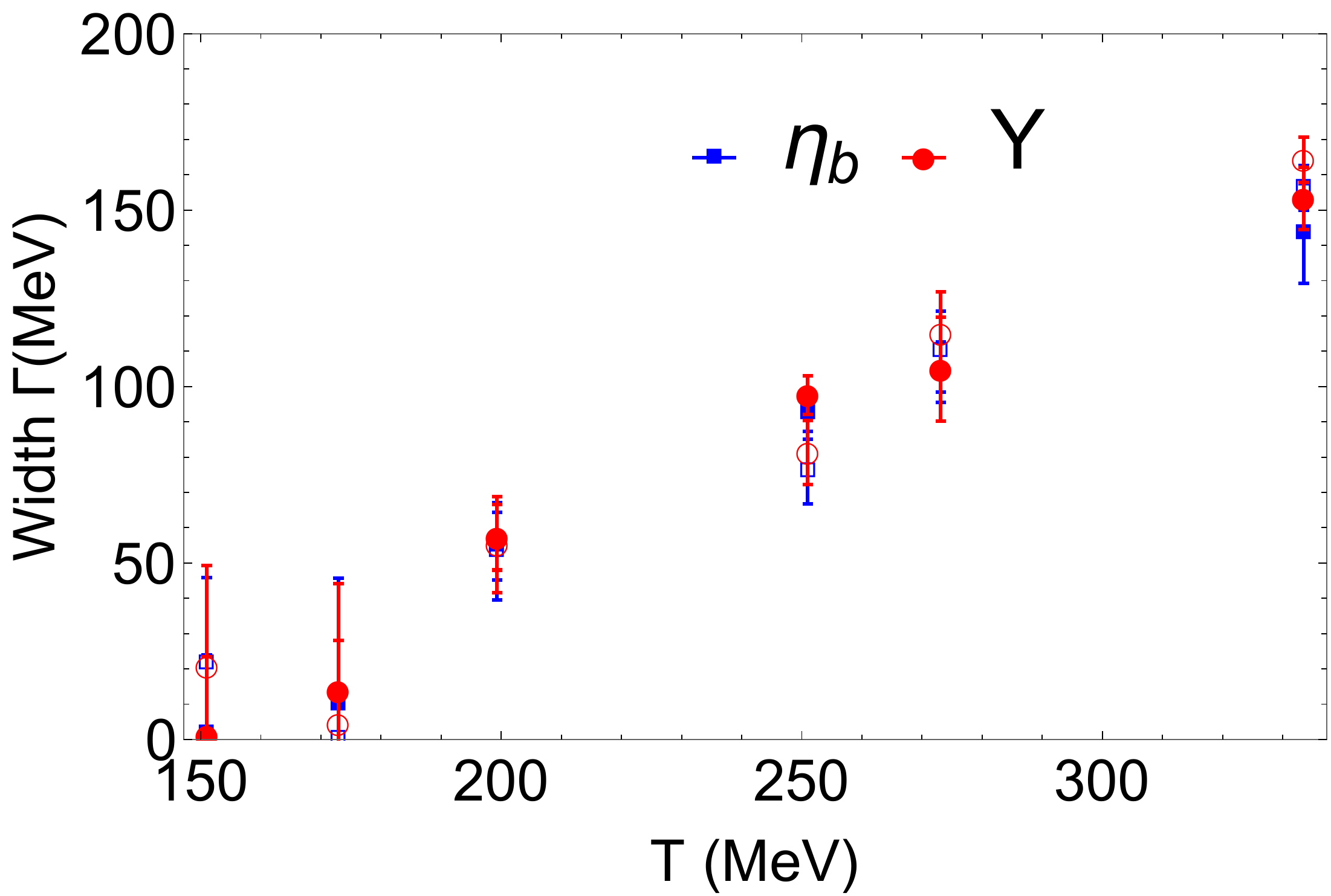}
\hspace{0.05\textwidth}
\includegraphics[width=0.45\textwidth]{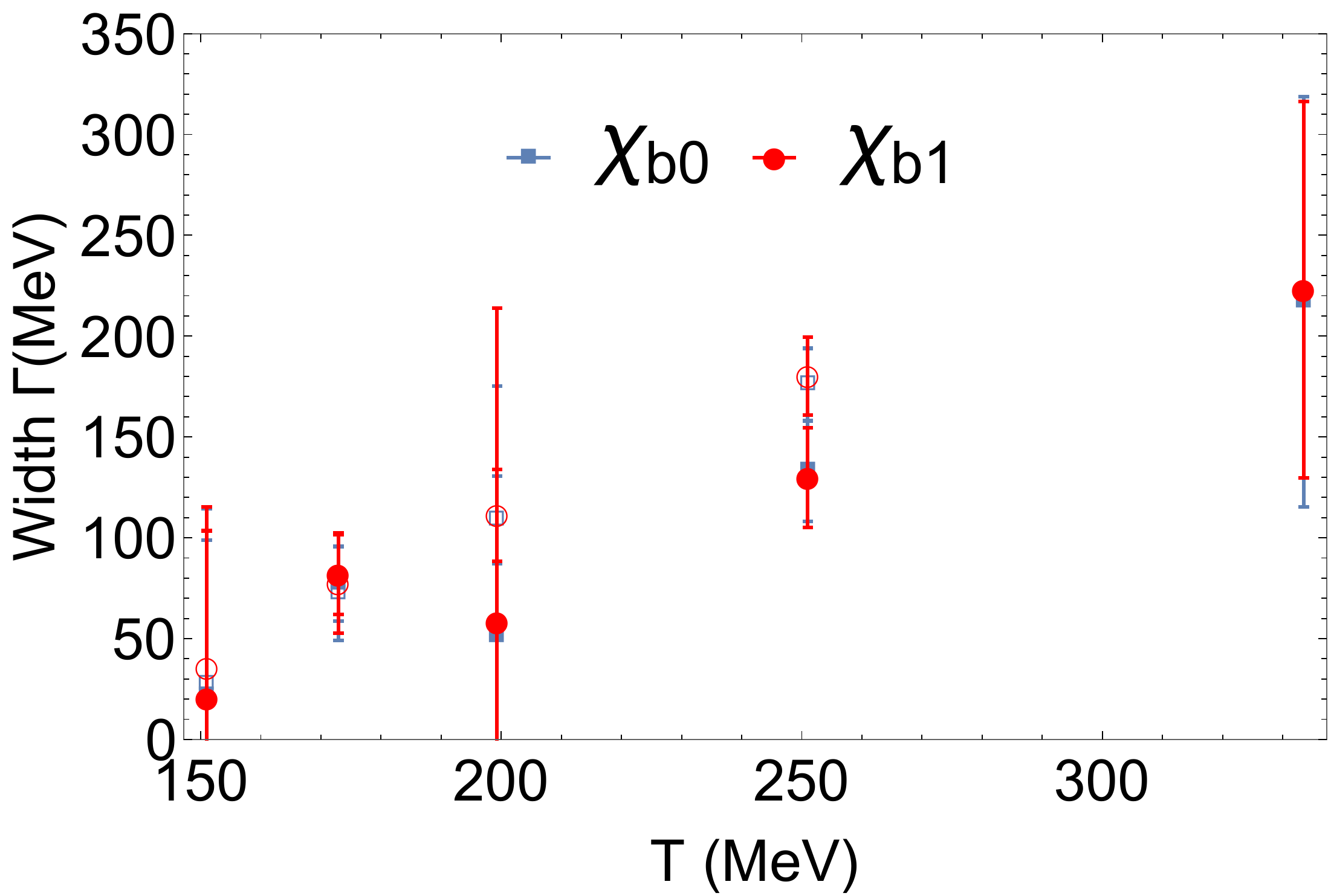}
\caption{In-medium width parameter, $\Gamma$, from fits to Eq.~(\ref{gauss}) for 1S
bottomonium (left) and 1P bottomonium (right). The open symbols correspond to fits
that exclude the data at the largest two $\tau$ values and $A_{cut}=0$, while the
filled symbols correspond to fits with all the data points. Open points does not work
for $\chi _b$ at the highest temperature and are thus not shown.}
\label{fig:GammaT}
\end{figure*}

As can be seen from Fig.~\ref{fig:dmeff_fit_gauss}, fits using Eq.~(\ref{gauss})
well-describe the lattice results on  $M_{eff,sub}(\tau,T)$ . The $\chi ^2/dof$ lies
in the range of 0.2 to 0.7 for $\eta_b$ and $\Upsilon$, and  0.1 to 0.6 for $\chi
_b$. From the fits we obtain the in-medium bottomonium mass $M(T)$ and the width
parameter $\Gamma(T)$. To test the robustness of the fit procedure, we also performed
fits omitting two data corresponding to the two largest $\tau$ values, and by setting
$A_{cut}=0$. Such fits work well for the effective
masses of $\eta_b$ and $\Upsilon$ at all temperatures as well as for $\chi_{b0}$ and
$\chi_{b1}$ except at the highest temperature. We began the fits from $\tau=a$ or $2a$, and used the average of the two
fit results; the difference of these two fit results were chosen as the systematic
errors. This systematic error was added in quadrature to the statistical error
calculated using jackknife sampling. Our results for the medium masses are shown in
Fig.~\ref{fig:massT} in terms of the mass differences (shifts) $\Delta
M(T)=M(T)-M(T=0)$ for 1S and 1P bottomonium states. We do not show results for $\chi_{b2}$ and $h_b$ since they are very similar to $\chi_{b0}$ and $\chi_{b1}$. The mass shift is compatible with
zero when statistical and systematic uncertainties are taken into account. The
in-medium width parameter for different bottomonium states is shown in
Fig.~\ref{fig:GammaT}. The fits generally work well, but some systematic errors does
appear, as seen in the right plot for $\chi _b$ at $T=199$~MeV. At this temperature
the linear behavior in the effective mass generated by the Gaussian form can equally
be well produced by the negative peak, which was introduced to explain the strong
drop off at $\tau = 1/T$. For the same reason in the fits for $\eta_b$ and
$\Upsilon$ at $T \leq 251$~MeV and for $\chi _b$ at $T\leq 173$~MeV, the negative
peak has not been included in the fit, since the data points appear to follow a linear
behavior to a very accuracy. The fit can not see the difference between a fit with a
Gaussian or a fit with two delta functions. We see from the figure that at the lowest
temperature the width is compatible with zero for both 1S and 1P bottomonia. The
width of $\eta_b$ and $\Upsilon$ is small for $T=173$`MeV, while it becomes
significant for $\chi_{b0}$ and $\chi_{b1}$. In general, the width is larger for 1P
bottomonia than for 1S bottomonia. This is expected since 1P bottomonia being larger
are more affected by the deconfined medium. The extracted values of the width
parameter are, in most cases, not sensitive to the simplified description of the low
$\omega$ tail, since the results from the fits using all data points agree mostly
with the results from the fits without the last two data points and $A_{cut}=0$.

\begin{figure*}[!t]
\centering
\includegraphics[width=0.45\textwidth]{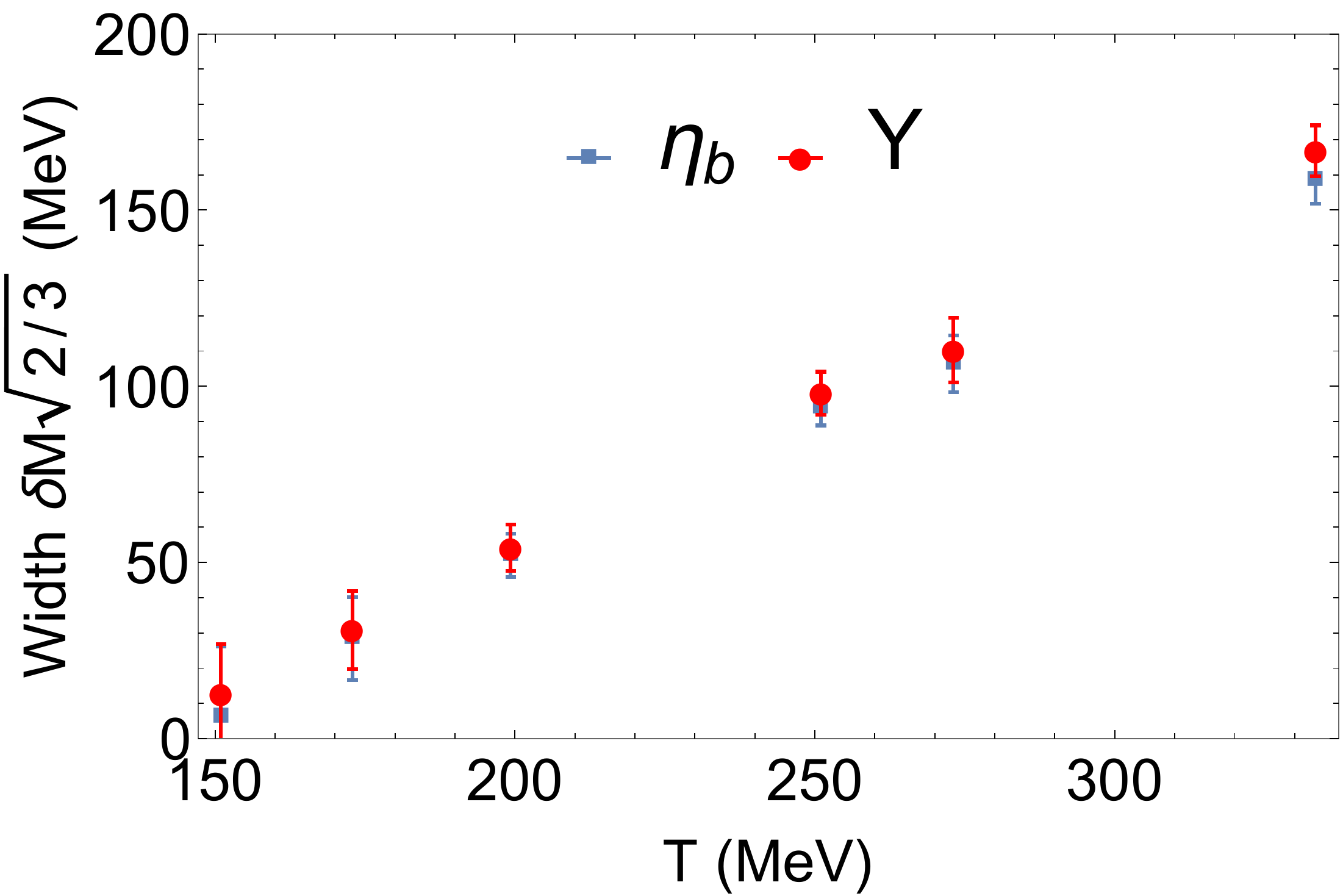}
\hspace{0.05\textwidth}
\includegraphics[width=0.45\textwidth]{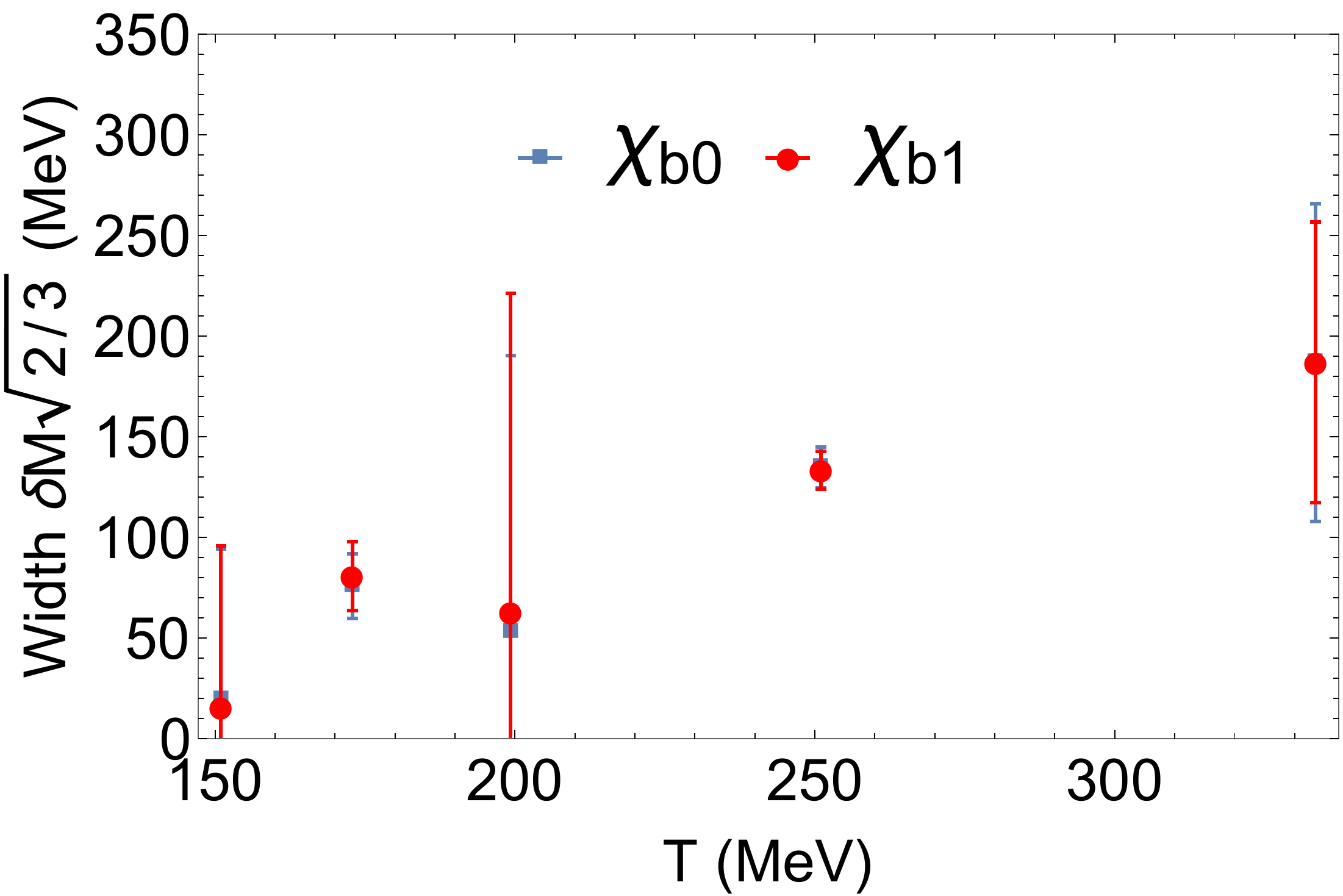}
\caption{In-medium width parameter $\delta M \sqrt{2/3}$ from fits to Eq.~
(\ref{3pk}) for 1S bottomonium (left) and 1P bottomonium (right). The fits are with
all the data points.}
\label{3peaks_width}
\end{figure*}

It is important to ask the question to what extent the above results for the
in-medium mass shift and width depend on the interpolating operator used in the
analysis. The continuum part of the spectral function clearly depends on the choice
of the interpolating operator. However, this part is, to a large extent, temperature
independent and does not show up in the subtracted effective masses. If the in-medium
bottomonium mass and width are physical then they should not depend on the choice of
the interpolating operator. On the other hand, the low energy tail of the spectral
function may depend on the choice of the interpolating operator. We discuss this
issue in Appendix~\ref{app:size_dep}, where we study the bottomonium correlators
corresponding to interpolating operators of different sizes. As shown in this
Appendix, the in-medium bottomonium mass and width only mildly depends on the size of
the interpolating operators, and this dependence is likely due to the imperfection of
the fit Ansatz of the spectral function.

The absence of a significant bottomonium mass shift at finite temperature should not
come as a complete surprise. The energy levels that enter the spectral decomposition of the bottomonium correlator are the same as in the vacuum, i.e., are not affected by the temperature, and include
the vacuum bottomonium state. This statement is not completely correct if NRQCD is used. However, as long as
the temperature is not too high and NRQCD is a good approximation, the energy levels
are about the same as at T=0.

 An effective in-medium shift of the peak position of
the spectral function can come from additional energy levels corresponding to
multi-particle states containing quarkonium. The density and distribution of these
energy levels could be such that they correspond to a peak of some width, with a
maximal density at an $\omega$ value that is different from the vacuum quarkonium
mass. We do not see any indication of such shift in our analysis. One may wonder if
this is due to the small volume used in our calculations, since the density of the
additional states could be too small and the location of the corresponding energy
levels could be highly distorted. Therefore, for the highest temperature $T=334$~MeV
we performed calculations on a smaller $36^3 \times 12$ lattice. The results are
discussed in Appendix~\ref{ap_Volume_effects}. As shown in this Appendix, we do not
see any significant volume effects in the bottomonium correlators, thus, it is not
clear if the absence of in-medium mass shift is due to finite volume effects.
Instead, it is possible that the absence of an effective mass shift is due to the use
of a simple Ansatz for $\rho_{med}$.

For finite volume the most natural representation of the in-medium spectral function
is the sum of $\delta$-functions. If we want to represent $\rho_{med}(\omega,T)$ by a
sum of $N$ $\delta$-functions, we have to fit $2N$ parameters and, thus, $N$ cannot
be too large. We have seen from the previous analysis that two to three parameters
are sufficient to describe the $\tau$-dependence of the subtracted effective masses,
which show an approximate linear behavior at small and moderate $\tau$ and a fast
drop-off at large $\tau$ that corresponds to the low $\omega$ part of
$\rho_{med}(\omega,T)$. We have also seen that bottomonium masses are not modified
with respect to their  vacuum values. The simplest representation of
$\rho_{med}(\omega,T)$ in terms of $\delta$-functions consistent with the above
features is
\begin{eqnarray}
\rho_{med}(\omega,T) &=& A_{cut}(T) \delta(\omega-\omega_{cut}(T)) \nonumber \\
&+& \delta(\omega-M_0+\delta M) \nonumber \\
&+& \delta(\omega-M_0) \nonumber\\
&+& \delta(\omega-M_0-\delta M)  \,,
\label{3pk}
\end{eqnarray}
where $M_0$ is the vacuum bottomonium mass, and the parameters $A_{cut}$ and
$\omega_{cut}$ describe the low $\omega$ tail of the spectral function. It is easy to
see that for small $\delta M$ the above Ansatz gives an effective mass that decreases
linearly in $\tau$, with the slope equal to $\delta M \sqrt{2/3}$ if the contribution
proportional to $A_{cut}$ can be neglected. Therefore, $\delta M \sqrt{2/3}$ can be
interpreted as a width parameter. We performed fits of the lattice results using
Eq.~(\ref{3pk}),  and obtained the values $\delta M$, $A_{cut}$ and $\omega_{cut}$.
Our results for $\delta M \sqrt{2/3}$ are shown in Fig.~\ref{3peaks_width}. We see
from the figure that the width parameters obtained from the fits with Eq.~(\ref{3pk})
agree well with the estimate of the bottomonium states obtained using the Gaussian
form of $\rho_{med}$. This suggest that our estimates of the width are robust.

\section{Conclusion}

We studied S- and P-wave bottomonium correlators at non-zero temperature using
extended Gaussian, as well as point meson sources within the framework lattice NRQCD
including $v^6$ spin-dependent terms. Correlators of point sources show little
temperature dependence because of their limited sensitivity to bottomonium properties
at small Euclidean time. Therefore, we focused on analyses of the correlators with
extended Gaussian meson sources. We identified the contributions of the high energy
part of the spectral functions to the correlators of extended Gaussian sources at
$T=0$. This contribution was then subtracted from the correlators at finite
temperature, significantly simplifying the analyses. The $\tau$-dependence of the
subtracted correlators with extended Gaussian sources could be understood in terms of
a simple theoretically-motivated spectral functions, consisting of a single broadened
peak. We identified two prominent structures in the corresponding effective masses:
an approximately linear decrease at small and intermediate values of $\tau$, which is
related to the width of the peak, and a rapid drop in the spectral function around
$\tau \simeq 1/T$, corresponding to the small $\omega$ tail of the peak. At present
statistical accuracy these features are very well described just by three
parameters.

We estimated the thermal width of bottomonium properties using two different forms of
the spectral functions that capture the above behavior of the effective masses. We
found that the thermal broadening of P-states is larger than the thermal broadening
of the S-states, and sets in at lower temperatures. This is expected based on the
difference in size of S-state and P-states. The in-medium bottomonium masses, defined
as peak positions, did not change relative to their vacuum values. This may appear
somewhat unexpected, especially for the P-states. We did not find any indications
that the absence of mass shift being due to finite volume of the lattice.  At current
level of statistical accuracy the lattice results of the correlators cannot resolve
further details on the shape of the spectral function beyond its overall width.
Therefore, to resolve the detailed shape of the spectral function, including its
asymmetric nature, and possible shift in the peak position, more precise lattice
results are needed. It is possible that with increase statistical precision one will
see more sensitivity to finite volume effects too.

\begin{acknowledgements}

  This material is based upon work supported by the U.S. Department of Energy, Office
  of Science, Office of Nuclear Physics: (i) Through the Contract No. DE-SC0012704;
  (ii) Through the Scientific Discovery through Advance Computing (ScIDAC) award
  Computing the Properties of Matter with Leadership Computing Resources. Stefan Meinel acknowledges support by the U.S. Department of Energy, Office of Science, Office of High Energy Physics under Award Number DE-SC0009913.

  This research used awards of computer time: (i) Provided by the USQCD consortium at
  its Fermi Natioanl Laboratory, Brookhaven National Laboratory and Jefferson
  Laboratory computing facilities; (ii) Provided by the INCITE program at Argonne
  Leadership Computing Facility, a U.S. Department of Energy Office of Science User
  Facility operated under Contract No. DE-AC02-06CH11357.

\end{acknowledgements}

\appendix

\section{Smearing Dependence} \label{app:size_dep}

\begin{figure}[!t]
 \centering
 \includegraphics[width=0.45\textwidth]{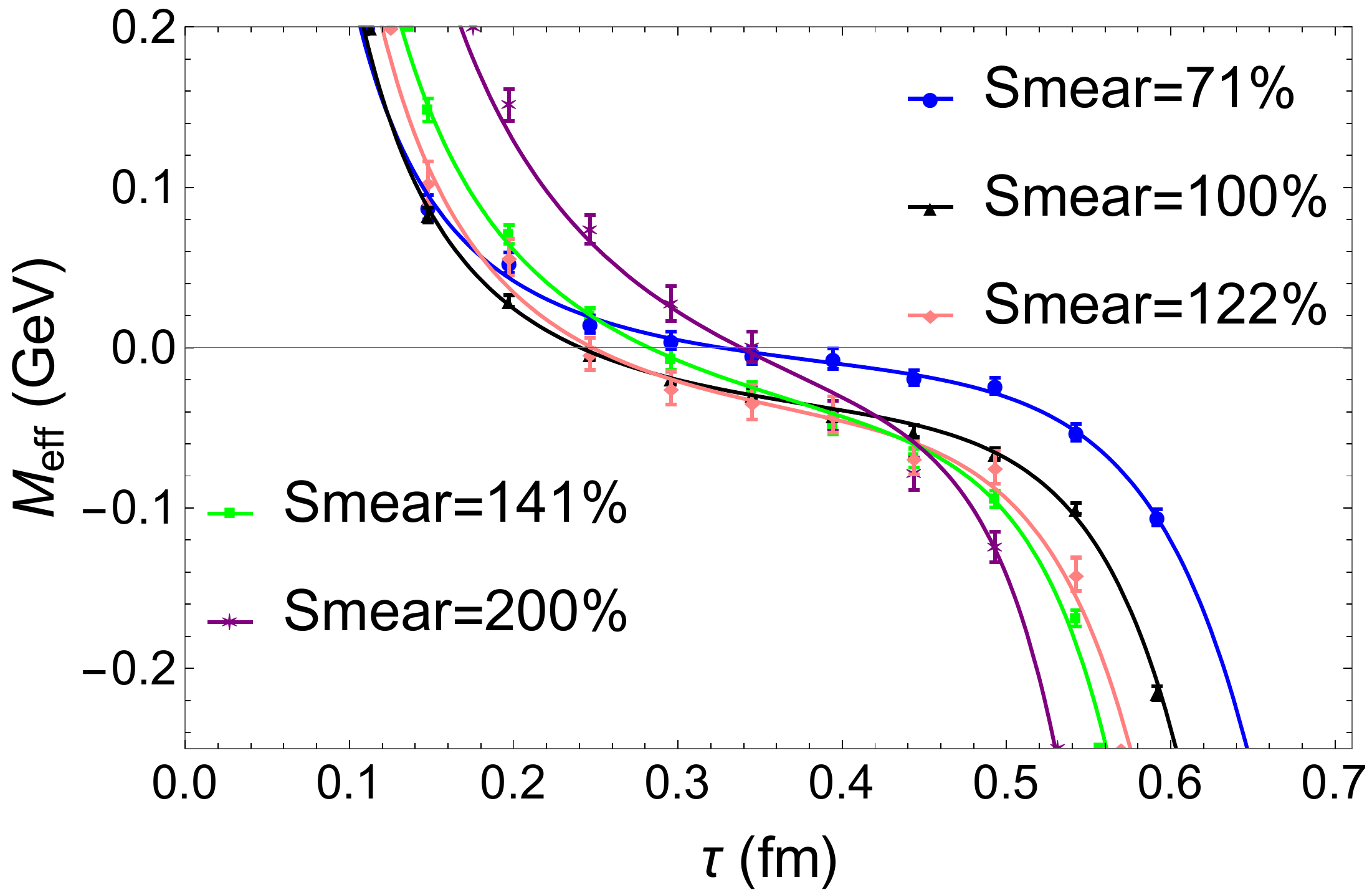}
   \caption{Effective mass for $\eta _b $ for $T=333.5$~MeV. The smeared sources
   correspond to size 0.15~fm (71\%, blue), 0.21~fm (100\%, black), 0.25~fm (122\%,
   pink), 0.29~fm (141\%, green) and 0.41~fm (200\%, purple). }
   \label{etab_smooth_com}
\end{figure}

In the main analysis we use Gaussian sources of size 0.21~fm, which we will refer
here to as the standard sources size. We studied to what extent our conclusions about
in-medium bottomonium properties depend on the choice of the source size. Therefore,
we calculated $\eta_b$ correlators at $T=334$~MeV using source sizes 0.15~fm,
0.21~fm, 0.25~fm and 0.41~fm, which are 71\%, 122\%, 141\% and 200\% of the standard
source size used in the main analysis, respectively. The corresponding numerical
results for the $\eta_b$ effective masses are shown in Fig.~\ref{etab_smooth_com}.
The qualitative behavior of the effective masses obtained for different smearing
sizes is similar, except for the largest sources size (200\% of the standard source
size), where no plateau like structure can been seen.

\begin{figure}[!t]
  \centering
  \includegraphics[width=0.45\textwidth]{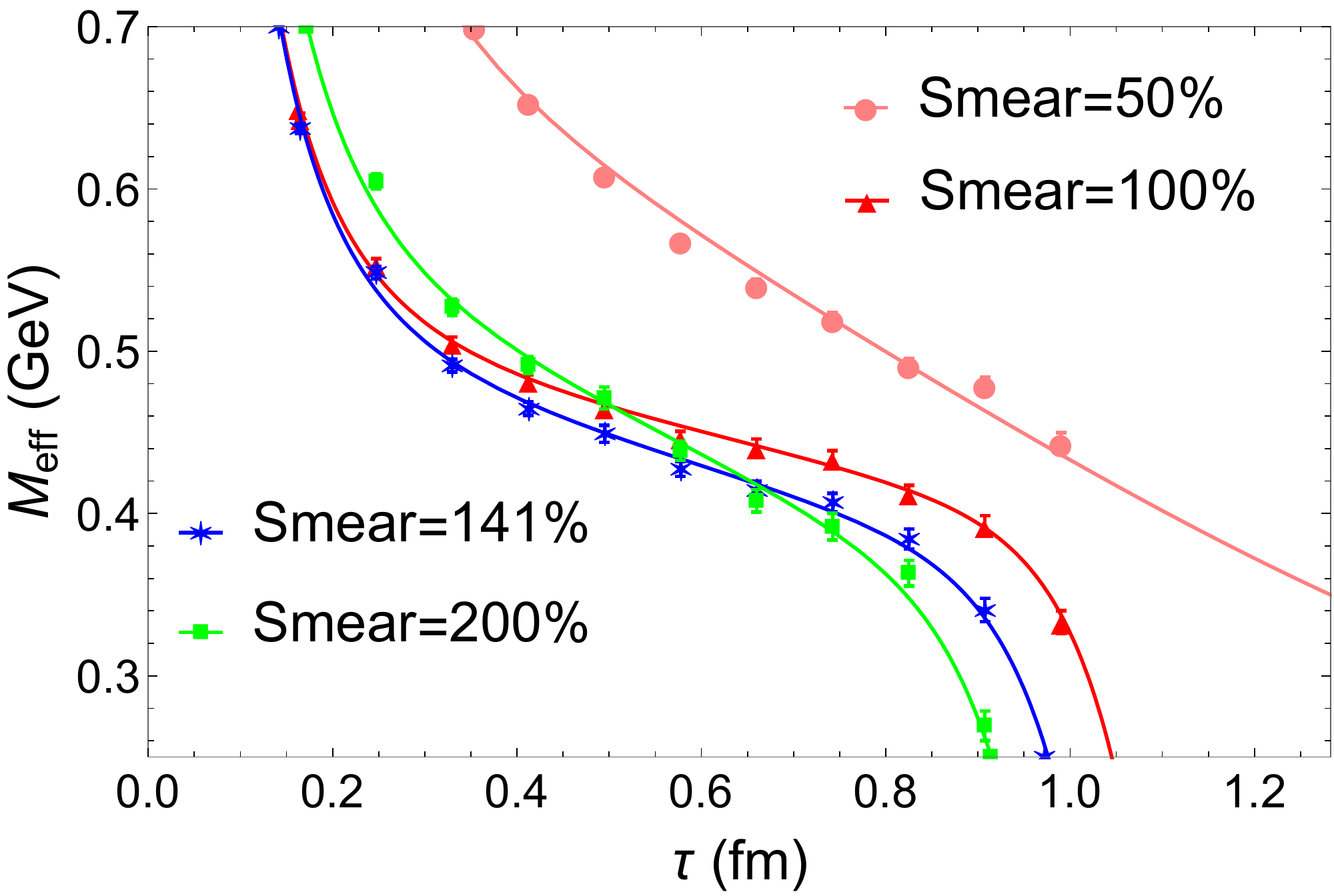}
   \caption{Effective mass for $\chi _{b0}$ for $T=199.3$~MeV. The smeared sources
   correspond to size 0.10~fm (50\%, pink), 0.21~fm (100\%, red), 0.29~fm (141\%,
   blue) and 0.41~fm (200\%, green). }
   \label{pwave_smooth_com}
\end{figure}

We also calculated the P-wave correlators at $T=199$~MeV for different source sizes,
namely for sources sizes that are 50\%, 141\% and 200\% of the standards source size.
The results for the effective masses are shown in Fig.~\ref{pwave_smooth_com}. We see
that the source size of 0.1~fm is too small, the corresponding effective mass is very
large and does not approach a plateau. We also see no clear plateau for the largest
source size corresponding to 200\% of the standard source size. The effective masses
obtained using source size of 0.21~fm and 0.29~fm look similar, but the effective
masses are smaller at small $\tau$ for the source size of 0.29~fm. Therefore, for
P-waves a better choice for the source size is around 0.29 fm.

To check the dependence of in-medium bottomonium properties on the smearing size, we
performed fits on the effective masses using the following Ansatz for the spectral
function
\begin{eqnarray}
\rho(\omega,T) &=& A_{cut} \delta(\omega-\omega_{cut}) \nonumber \\
&+& A \exp \left[ -\frac{(\omega-M)^2}{2 \Gamma^2} \right] \nonumber \\
&+& B \theta(\omega-\omega_0) \,.
\label{rho_all}
\end{eqnarray}
The last term in this equation parameterize the continuum part of the spectral
function. We performed fits on the effective masses obtained with source sizes of
71\% to 141\% of the standard source size and extracted the in-medium masses and
width of the bottomonium states. The in-medium parameters obtained from the effective
masses with different source sizes are similar but not exactly the same. This is due
to the fact that here we fit the continuum part of the spectral function, which is
different for different source sizes. The continuum parameters are correlated with
the mass and width parameter and thus the latter are also affected. It is possible
that the simple fit form given by Eq. ~(\ref{rho_all}) is less appropriate for source
sizes that are not around the optimal value.

\begin{figure}[!t]
  \centering
  \includegraphics[width=0.45\textwidth]{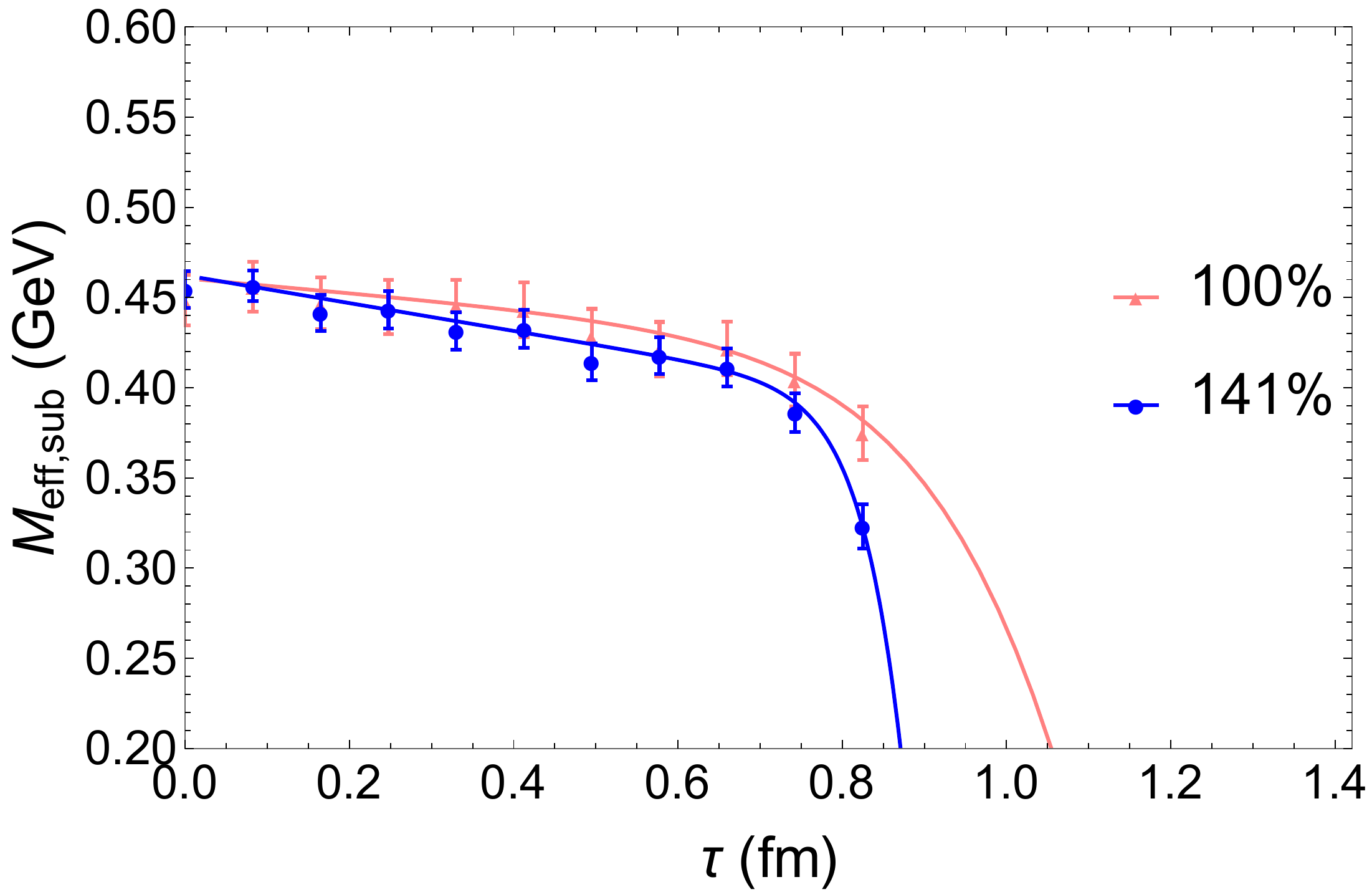}
   \caption{Effective mass for $\chi _{b0}$ for $T=199.3$~MeV. Smeared source size
   0.21~fm (100\%, pink), 0.29~fm (141\%, blue). Zero temperature continuum
   subtracted.}
   \label{smearing_100_vs_144_sub}
\end{figure}

In order to have a more robust comparison of the results obtained with different
source sizes, for the source size of 0.29~fm we also performed calculations of the
P-wave correlator at zero temperature. This enabled us to calculate the high energy
part of the correlator, $C_{high}(\tau)$, and subtract it from the corresponding
$T>0$ result. The resulting subtracted effective mass is shown in
Fig.~\ref{smearing_100_vs_144_sub}, and compared to the result obtained with source
size of 0.21~fm discussed in the main text. The subtracted effective masses for these
two source sizes appear to be quite similar, though the effective mass corresponding
to source size of 0.29~fm shows a faster fall-off at large $\tau$. This means that
the low $\omega$ tail is more important for the larger source size. We performed fits
on the subtracted effective masses for the above two source sizes using
Eq.~(\ref{gauss}) for the spectral function, and obtained the in-medium mass and
width. We find that the in-medium mass agrees with the vacuum mass also for the
source size of 0.29~fm. For the in-medium width parameter we obtain: $\Gamma=86 \pm
55$~MeV for the standard source size and $\Gamma=123\pm 23$~MeV for the source size of
0.29~fm. The width parameter is $36 \pm 59$~MeV larger for the larger source size.
This is likely due to the correlation between the Gaussian part of the spectral
function and the low $\omega$ tail.

\section{Subtracted zero temperature correlator} \label{ap_subtracted_cor}

\begin{figure*}[!t]
\centering
\includegraphics[width=0.45\textwidth]{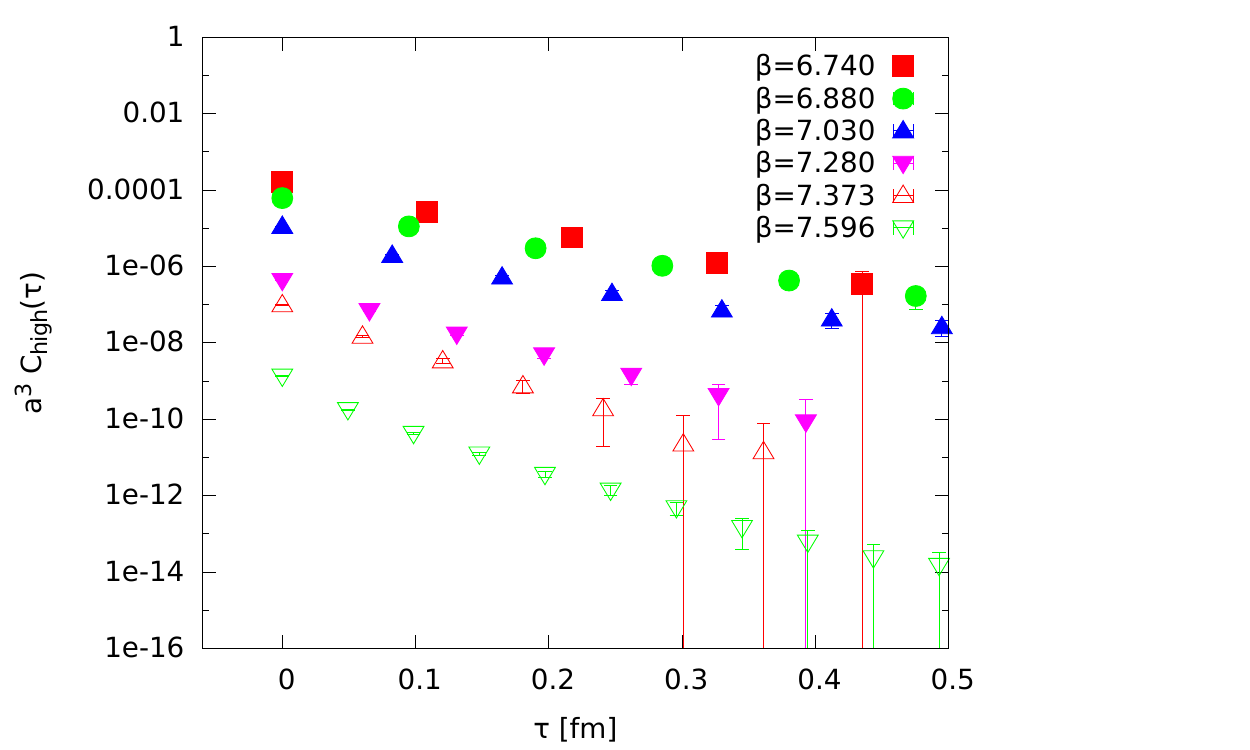}
\hspace{0.05\textwidth}
\includegraphics[width=0.45\textwidth]{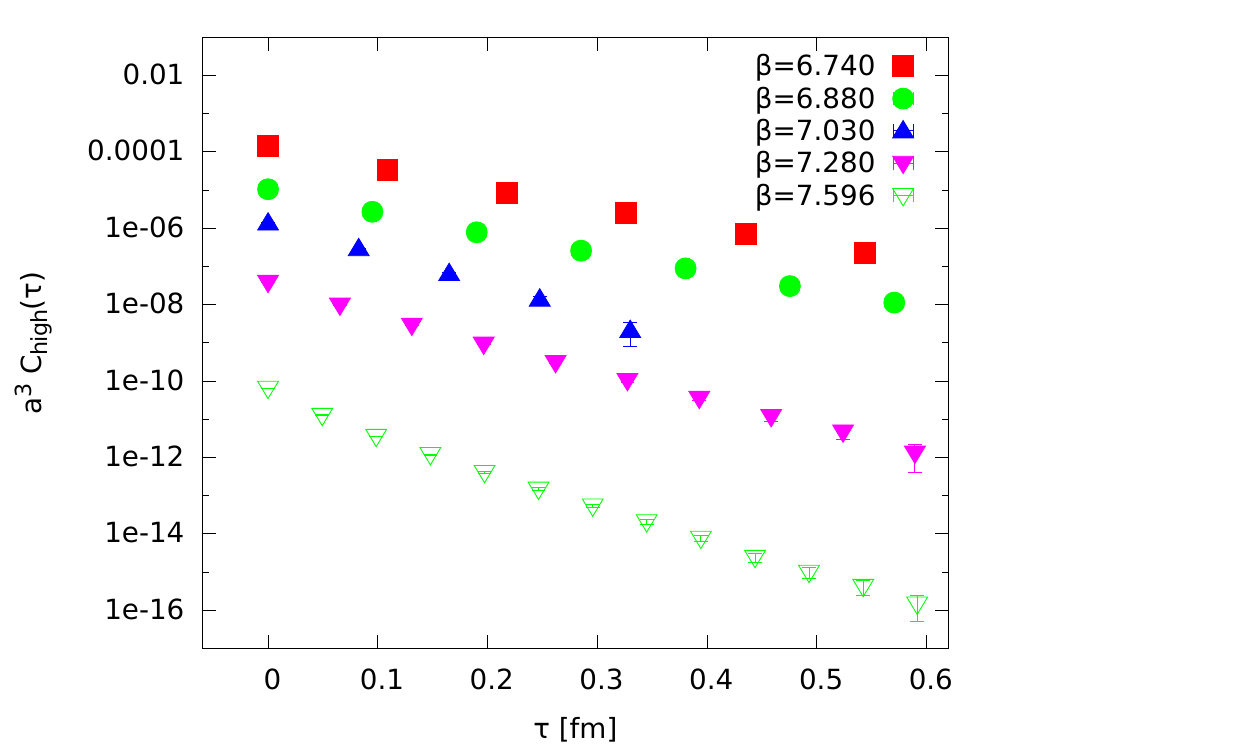}
\caption{The high correlator $C_{high}(\tau)$ for $\eta_b$ (left) and $\chi_{b0}$
(right) as function of $\tau$. }
\label{fig:Chigh}
\end{figure*}

The $T=0$ correlator with the ground state subtracted, $C_{high}(\tau)$, is shown in
Fig.~\ref{fig:Chigh} for $\eta_b$ and $\chi_{b0}$ at different values of $\beta$.
Following Eq.~(\ref{CT0}), a fit to the  correlator on the part that is a plateau
when plotting the effective mass is first performed using an exponential function.
This exponential function is then subtracted from the entire correlator. This leaves
only the high $\omega$ part left. This procedure is expected to work only because the
state we are looking for dominates the contribution, due to the smearing. At large $\tau$ the relative errors on
$C_{high}$  are very large because the correlator is dominated by the ground state.
However, for small tau we can determine $C_{high}$ reliably.

\section{Volume effects}\label{ap_Volume_effects}

\begin{figure*}[!t]
  \centering
  \includegraphics[width=0.45\textwidth]{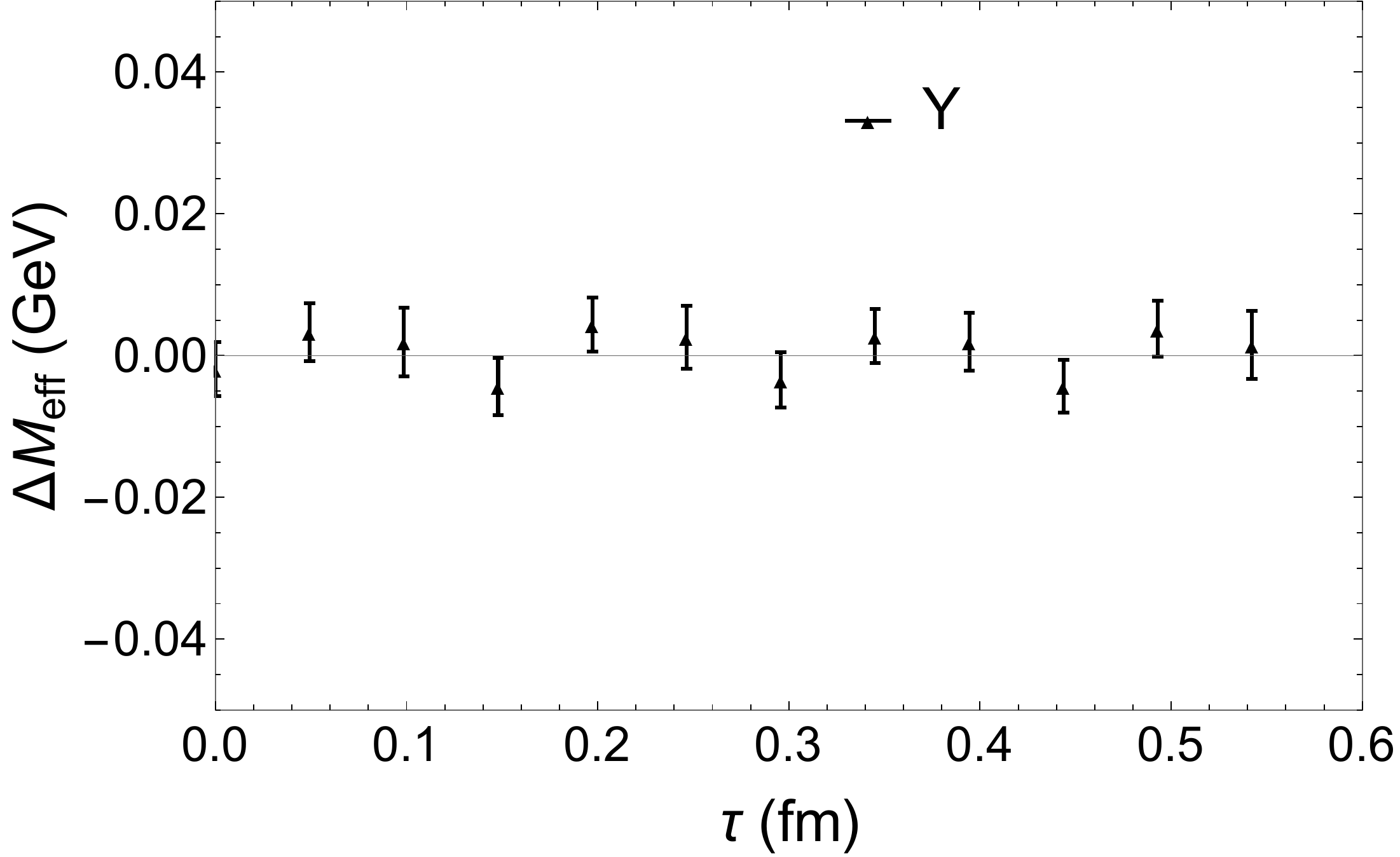}
  \hspace{0.05\textwidth}
  \includegraphics[width=0.45\textwidth]{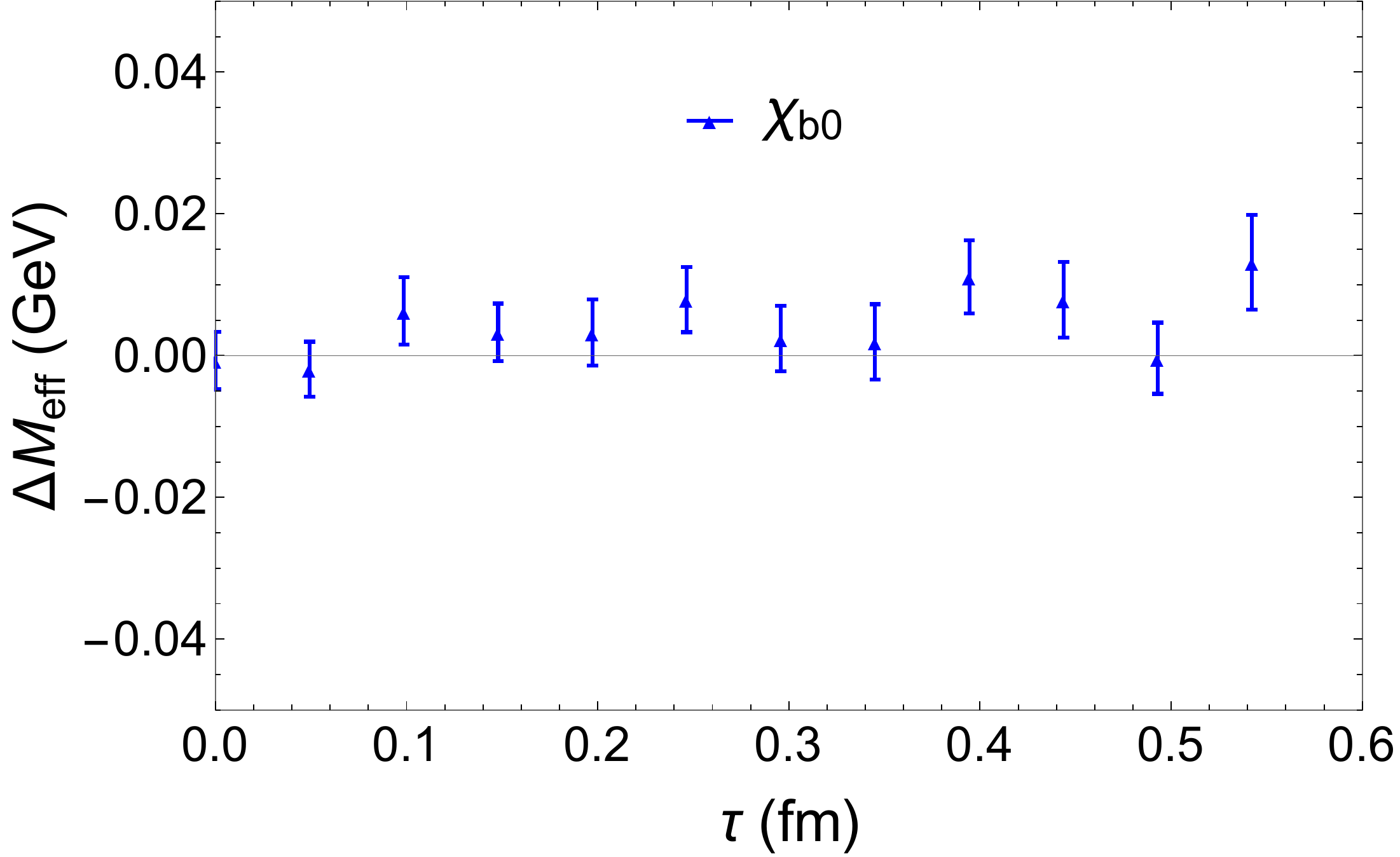}
   \caption{The difference of the the effective masses calculated on $36^3 \times 12$
   lattice and $48^3 \times 12$ lattice for $\Upsilon$ (left) and $\chi_{b0}$ (right)
   at $T=334$~MeV.}
   \label{fig_vol_effect}
\end{figure*}

The in-medium modifications of the spectral function may be sensitive to the volume,
since in finite volume the spectral function is given by sum of $\delta$-functions
and the number of term in this sum is volume dependent. Furthermore, the energy
levels that enter the $\delta$-functions are also volume dependent. To explore finite
volume effects we performed additional calculations of bottomonium correlators at the
highest temperature $T=334$~MeV using $36^3 \times 12$ lattice, 1344 gauge
configurations and 96 sources per gauge configuration. The effective masses obtained
from these calculations turned out to be very similar to the ones obtained on $48^3
\times 12$ lattice. In fact, the largest deviation we have seen between the effective
masses obtained on two different lattice volumes was $1.5\sigma$. As an example in
Fig.~\ref{fig_vol_effect} we show the difference of the effective masses calculated
on $36^3 \times 12$ lattice and $48^3 \times 12$ lattice as function of $\tau$ for
$\Upsilon$ and $\chi_{b0}$. At present statistical accuracy we do not see volume
dependence of quarkonium correlators at the highest temperature.

\bibliography{ref}

\end{document}